# MASTER THESIS

## Structure and Properties of Thermoresponsive Diblock Copolymers Embedded with Metal Oxide Nanoparticles

LUDWIG-MAXIMILIANS-UNIVERSITÄT MÜNCHEN

TECHNISCHE UNIVERSITÄT MÜNCHEN

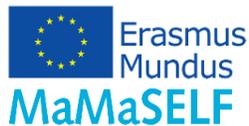 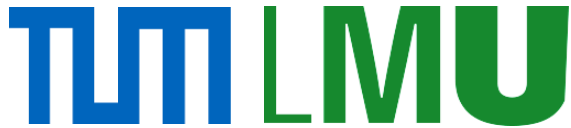 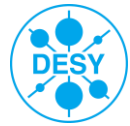 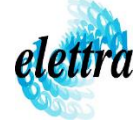 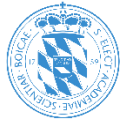

*Author:*
**Mr. Hong XU**
TUM/LMU
hong.xu@ph.tum.de

*Supervisor:*
**Prof. Dr. Peter MÜLLER-BUSCHBAUM**
Physik-Department E13, TUM
muellerb@ph.tum.de

*Examiner:*
**Prof. Dr. Wolfgang SCHMAHL**
Sektion Kristallographie, LMU
wolfgang.w.schmahl@lrz.uni-muenchen.de

*Advisor:*
**Dr. Ezzeldin METWALLI**
Physik-Department E13, TUM
ezzmet@ph.tum.de

"MASTER THESIS SUBMITTED TO THE FACULTY OF EARTH- AND ENVIRONMENTAL SCIENCES OF LUDWIG-MAXIMILIANS-UNIVERSITÄT MÜNCHEN IN THE FRAMEWORK OF ERASMUS MUNDUS MAMASELF"

September 30$^{th}$ 2016

# Abstract

Nanostructured polymer-metal oxide composites are a current research area of great importance due to its highlight applications in sensors, optics, catalysts and drug delivery [1-8]. Particularly the use of thermoresponsive polymers gives more flexibilities and possibilities in the design and construction of polymer templates. In the present investigation, the structure and magnetic properties of hybrid metal oxide/DBC films composed of two kinds of polystyrene-block-poly(N-isopropylacrylamide) (PS-b-PNIPAM) diblock copolymers (DBCs) with PS and PNIPAM as the major polymer domains respectively, and iron oxide were investigated. The thermoresponsive PNIPAM has a lower critical solution temperature (LCST) in aqueous solution at $32^{o}C$ [9, 16], which enables the controllable volume ratio of PS and PNIPAM in the structure of PS-b-PNIPAM diblock copolymers (DBCs). Thus, a temperature and humidity controlling cell was designed and built for precisely tuning the block structure of PS-b-PNIPAM DBCs, which was investigated by in-situ small-angle X-ray scattering (SAXS) and grazing-incidence small-angle X-ray scattering (GISAXS) measurements. The superparamagnetic behavior of the heat-treated hybrid iron oxide/ PS-b-PNIPAM DBC films was investigated using a superconducting quantum interference device (SQUID) magnetometer.

# Contents

# List of Symbols and Abbreviations

$D$ ………………….. periodic domain distance

$H_c$ …………………. coercivity

$M_r$ ………………… remanence

$M_s$ ………………… saturation magnetization

$N$ …………………...degree of polymerization

$T_g$ …………………. glass transition temperature

$\chi$ ……………………Flory-Huggins parameter

$f$ ……………………volume fraction

BC …………………block copolymer

CYL ………………..cylinder

DBC ………………..diblock copolymer

DC ………………....direct current

DESY ……………....Deutsches Elektronen-Synchrotron

DI water ……………deionized water

FWHM ……………..full width at half maximum

GISAXS ……………grazing incidence small angle X-ray scattering

GYR……………….. gyroid

$H_2SO_4$ …………….. sulfuric acid

LAM ……………….. lamellar

MINAXS ………….. Micro- and Nanofocus X-ray Scattering

NIPAM ……………..N-isopropyl acrylamide

NP ………………….. nanoparticle

ODT ………………….order disorder transition

PNIPAM …………….poly(N-isopropyl acrylamide)

PS …………………….polystyrene

PS-b-PNIPAM ………polystyrene-block-poly(N-isopropyl acrylamide)

PS-b-PNIPAM ………PS dominated PS-b-PNIPAM

PS-b-PNIPAM ………PNIPAM dominated PS-b-PNIPAM

RF …………………. .radio frequency

r.H. …………………..relative humidity

SAXS ………………...small angle X-ray scattering

SEM …………………scanning electron microscopy

SLD …………………..scattering length density

SQUID ………………superconducting quantum interference device

THF …………………tetrahydrofuran

TUM…………………Technical University of Munich

# Chapter 1

## Introduction

Nanostructured polymer-metal oxide composites are a current research area of great importance due to its various domain structures with specific optical, thermal, electric mechanical or magnetic properties. Particularly the use of thermoresponsive diblock copolymers (DBC) gives more flexibilities and possibilities in the design and construction of polymer templates. By introducing in-corporate functional nanoparticles (NPs) into the diblock copolymer matrix, the hybrid diblock copolymer-metal oxide nanocomposite can be functionalized for potentially tremendous applications, such as sensors, optics, solar cells, thermoplastic resins, catalysts, drug delivery, photonic band gap materials and magnetic materials [1-8].

It has been widely investigated that the thermoresponsive poly(N-isopropylacrylamide) (PNIPAM) and its block copolymers in aqueous solution [9-10] and gel [11-12] system. The swelling and deswelling thermoresponsive behaviors of PNIPAM solution or gel are demonstrated in the previous studies. But it has been rarely studied for the thermoresponsive PNIPAM DBC in film or bulk form, which is significant and practical for very common applications as film or bulk materials. Y. Yao et al. investigated the arrangement of maghemite nanoparticles in polystyrene-block-poly(N-isopropylacrylamide (PS-*b*-PNIPAM) DBC films by wet chemical self-assembly method. And the change of the structure was probed by ex-situ grazing-incidence small-angle X-ray scattering (GISAXS) [13]. S. Schmidt et al. studied adhesion and mechanical properties of PNIPAM micro-gel film [14]. Vincent P. Gilcreest et al. investigated contact angles and surface energies of thermoresponsive PNIPAM copolymers [15]. Although there are some relative studies, none of them ever did the in-situ small angle X-ray scattering (SAXS) studies for metal oxide / PS-*b*-PNIPAM DBC hybrid nanocomposite bulks and films. The work in this thesis is the first time exploring the thermoresponsive properties for PS-*b*-PNIPAM DBC bulks and films using in-situ SAXS techniques.

In the present investigation, the structure and properties of hybrid bulks and films composed of two kinds of PS-*b*-PNIPAM DBCs with PS and PNIPAM as the major polymer domains respectively, and iron salt or iron nanoparticles were investigated. The PS-*b*-PNIPAM, as any other diblock copolymer, has its block structures various from sphere, gyroid, cylinder to lamellar corresponding to the initial volume ratio range of two blocks. While one of the block - thermoresponsive PNIPAM, has a lower critical solution temperature (LCST) in aqueous solution at 32℃° [9, 16], meaning that the PNIPAM block collapses above 32℃° with a great volume shrinkage and vice versa. Besides, the swelling/de-swelling of PNIPAM also limited by the PS block which is rigid in humidity circumstance regarding temperature change. Thus two kinds of PS-b-PINIPAM DBCs with either PS or PNIPAM as a major block were investigated.

To precisely control and monitor this thermoresponsive behavior of PNIPAM based BC, a temperature and humidity controlling cell was designed and built for tuning the block structure of PS-*b*-PNIPAM DBCs, which was installed in the Ganesha SAXSLAB in-house SAXS instrument at the physics department of Technical University of Munich (TUM). The controlling cell consists of temperature regulation system using Julabo water thermostat and humidity regulation system using ProUmid MHG100 modular humidity generator. A free standing film is mounted in a cell vertically allowing the X-ray beam passing through the film. The whole system was tested and operated continually long time for checking the system stability and compatibility.

Additionally, the structures of bulk metal oxide/PS-*b*-PNIPAM DBCs hybrid nanocomposites with different contents of either iron salt or PS-tailored iron oxide nanoparticles were probed by ex-situ SAXS measurements. The surface morphologies of pure PS-*b*-PNIPAM DBCs were probed by SEM measurements. The structures of the corresponding metal oxide/PS-*b*-PNIPAM DBCs hybrid film were also investigated. In-situ real time GISAXS data was collected during gold metal deposition on some selected PS-*b*-PNIPAM DBCs films. Finally, superparamagnetic behaviors of the heat-treated metal oxide/PS-*b*-PNIPAM DBCs hybrid nanocomposite films were investigated using a superconducting quantum interference device magnetometer (SQUID).

# Chapter 2

## Theoretical Aspects

### 2.1 Polymers

Polymers, also known as macromolecules, are built up of a large number of molecular units that are linked together by covalent bonds. This is defined by Prof. Gert Strobl in the book *Physics of Polymer (3$^{rd}$ Edition)* [17]. The definition also coincides the original interpretation of the ancient Greek word πολύς *(polus)* meaning *"many, much"* and μέρος *(meros)* meaning *"parts"* [18]. Generally, polymers range from daily used synthetic plastics like polyethylene to natural biopolymers for instance DNA and proteins, and both of them are created via polymerization of many small molecules, known as *monomers* [19].

#### 2.1.1 Polymer structure

The entire structure of a polymer is generated during polymerization, the process by which elementary units (chemical monomers) are covalently bonded together. The degree of polymerization N is the number of monomers in a polymer molecule [17]. And the molar mass M of a polymer is equal to its degree of polymerization $N$ times the molar mass $M_{mon}$ of its chemical monomer [20].

$$M = N M_{mon} \tag{2-1}$$

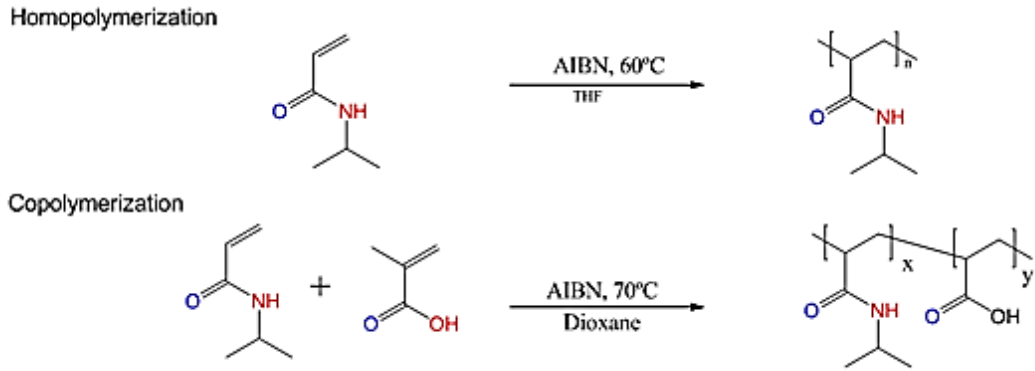

Figure 2-1: Homopolymerization of NIPAM and copolymerization of NIPAM and carboxylic acid [21].

There are many kind of polymers which usually can be classified into homopolymer or heteropolymer according to its constitutions. A homopolymer is a macromolecule that contains only one type monomer. It is made by the same monomer but could have different microstructure, degree of polymerization or architecture [20].

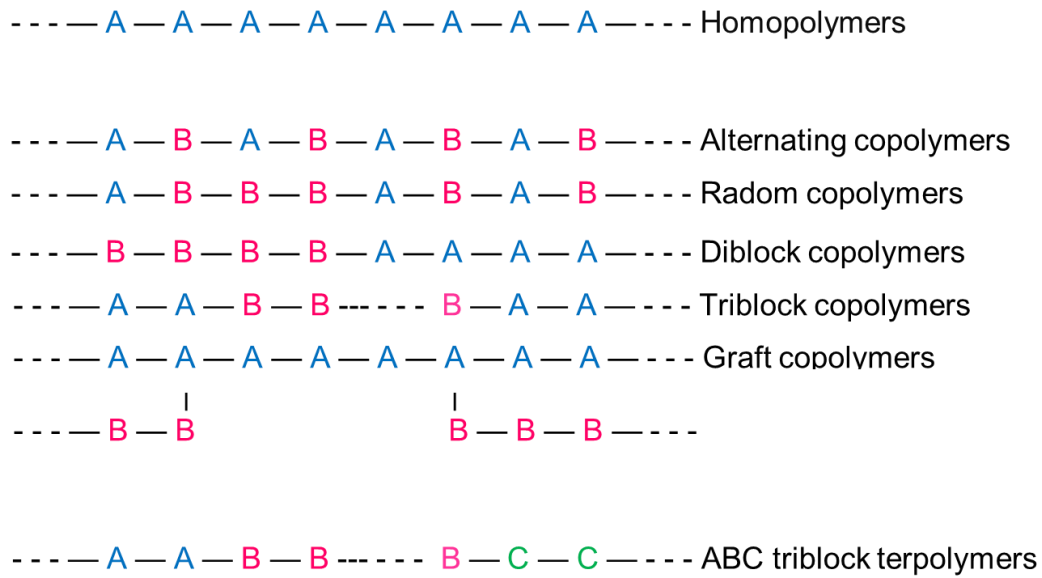

Figure 2-2: Structure of homopolymer and types of copolymers including alternating, random, diblock, triblock and graft copolymers, and ABC triblock terpolymers [20].

If several different types of monomers are combined into a single chain, this gives a new macromolecule called heteropolymer. Macromolecules containing two different monomers are called copolymers, which can be alternating, random, block or graft depending on the sequence of monomers. Especially for block copolymer, which is named as diblock copolymers if there are two blocks, triblock copolymers if there are three blocks, multiblock copolymers if there are many alternating blocks. In addition, polymers containing three types of monomers are called terpolymers, which is shown in figure 2-2 as an example of ABC triblock terpolymers [20].

The polymer network is also important for the properties of polymeric system. There are several types of polymer architectures including linear, star-branched, H-branched, comb, ladder, dendrimer or randomly branched as sketched in figure 2-3. The physical properties of a polymer like hardness, strength, flexibility and so on depend on the polymer architectures. [20]

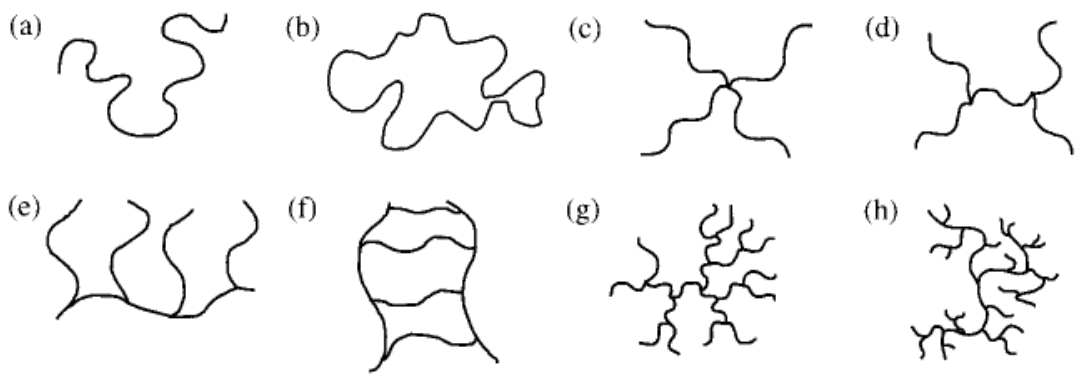

Figure 2-3: Example of polymer architectures: (a) linear; (b) ring; (c) star; (d) H; (e) comb; (f) ladder; (g) dendrimer; (h) randomly branched. [20]

## 2.1.2 Phase separation

For block copolymers, a phenomenon called microphase separation happens during the energetical treatment like heating or cooling. Because of the long relaxation time and large scale of polymer molecules, many special features like volume shrinking and phase inversion can be observed in the polymer systems [17]. Since the two blocks are linked by covalent bond, it is not possible for a block copolymer to separate macroscopically. But a separation of two blocks can happen.

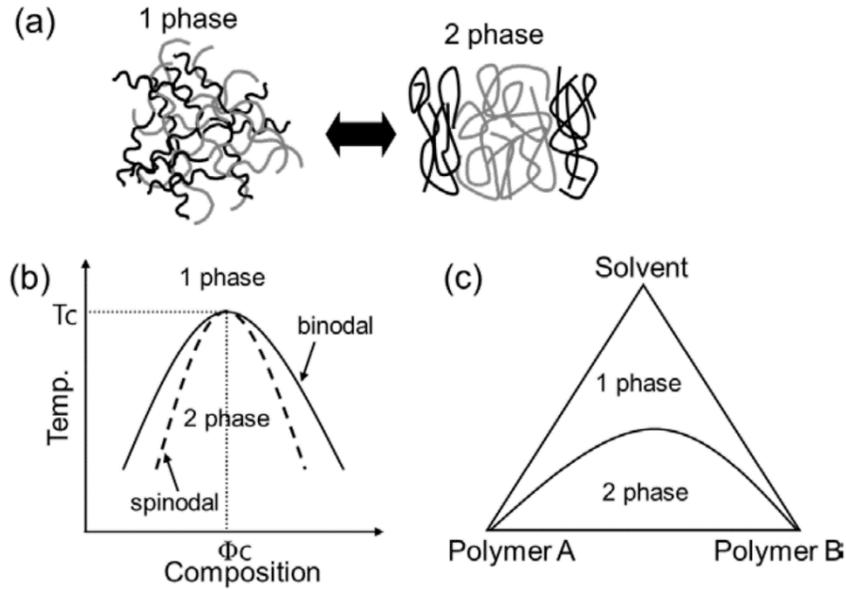

Figure 2-4: (a) Phase separation characteristics of a binary polymer blend; (b) Spinodal and binodal lines of a binary polymer blend as a function of composition and temperature; (c) A ternary phase diagram for the system with polymer A, polymer B, and the solvent [22].

The phase separation happens when the system physical condition changes. As an example shown in figure 2-4 (b) and (c), the phase diagrams describe the phase separations due to the temperature and solvent changes. To describe the phase separation behaviors of block polymers, a mean-field theory regarding polymer blends was introduced by Huggins and later was developed by Flory [23]. For a blend of two polymers, the Gibbs free energy is the most important relationship governing the mixed system:

$$\Delta G_m = \Delta H_m - T\Delta S_m \qquad (2\text{-}2)$$

Where $\Delta G_m$ is the free energy of mixing, $\Delta H_m$ is the enthalpy of mixing (heat of mixing) and $\Delta S_m$ is the entropy of mixing [23].

The entropy of the system can be calculated monomer by monomer, thus it gives the entropy $\Delta S_m$:

$$\Delta S_m = -RV \left( \frac{\varphi_1}{N_1} In\varphi_1 + \frac{\varphi_2}{N_2} In\varphi_2 \right) \qquad (2\text{-}3)$$

A parameter termed the Flory-Huggins interaction parameter, $\chi_{12}$, has been defined as: [23]

$$\chi_{12} = \frac{zw_{12}}{RT} \tag{2-4}$$

Where $z$ is the coordination number, $w_{12}$ is the exchange energy of interacting segments. This leads to the calculation of enthalpy of mixing: [23]

$$\Delta H_m = \varphi_1 \varphi_2 RTV \frac{\chi_{12}}{v_r} \tag{2-5}$$

Where $v_r$ can represent molecular or molar segment volumes. Thus the key equation of Flory-Huggins theory can be derived from the above calculation of the entropy and enthalpy of mixing.

$$\Delta G_m = RTV \left[ \frac{\varphi_1}{v_1} In\varphi_1 + \frac{\varphi_2}{v_2} In\varphi_2 \right] + \varphi_1 \varphi_2 \chi_{12} \frac{RTV}{v_r} \tag{2-6}$$

Where $v_i$ = molar volume of polymer chains $i$, $v_r$ = molecular or molar volume of a specific segment, which is also usually calculated as the square root of the product of the individual segmental unit molecular or molar volumes of the polymeric components. [23]

The temperature dependence of $\chi_{12}$ is often been expressed by:

$$\chi_{12} = a + \frac{b}{k_B T} \tag{2-7}$$

Thus the microphase separation can be described by Flory-Huggins theory together with the degree of polymerization degree $N$ and the volume faction of one block. [23]

When $\chi N \approx 10.5$, it is considered as weak segregation. When $10.5 < \chi N < 100$, it is intermediate segregation and the composition profile is sinusoidal about the mean value. When $\chi N > 100$, it is very strong segregation and the block domains are consist of pure components [24].

## 2.1.3 Diblock copolymers

As introduced in section 2.1.1 polymer basics, the diblock copolymer is a copolymer made up of two different blocks of polymerized monomers. The repulsive interaction between the different blocks leads them to separate while the strong chemical bond prevents them from a macro-phase separation and demixing. Finally, the diblock copolymers occurs to be separated in a mesoscopic scale, which is effected by the volume ratio of the blocks. Thus it can form a periodic ordered mesoscopic structures like lamellar, gyroid, cylinder and sphere as presented in figure 2-5.

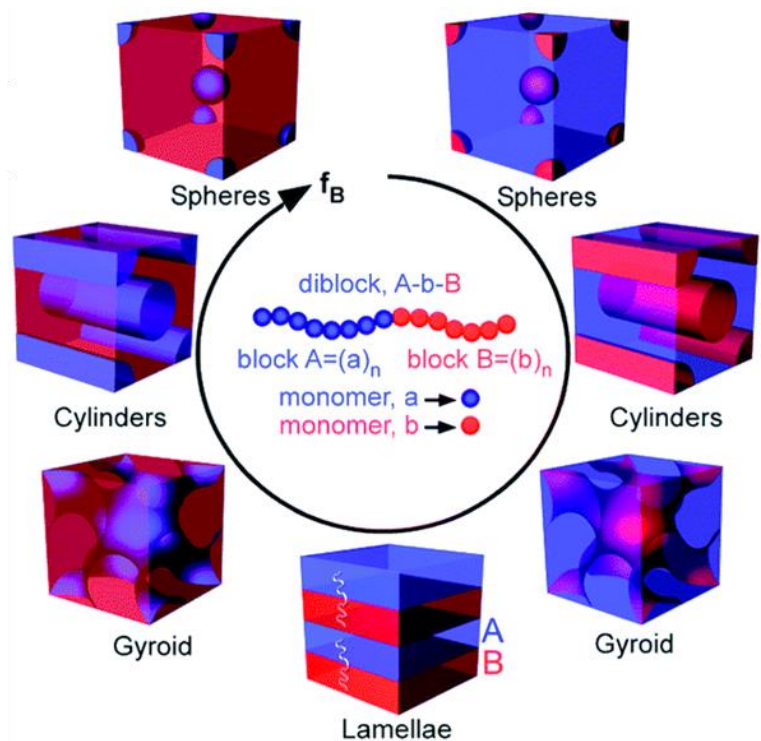

Figure 2-5: Illustrations of the typical equilibrium morphologies created via microphase separation in DBC systems with the increase of the volume fraction of the block B [25].

Due to different volume ratio of the polymer blocks, the diblock copolymer structure various from sphere, cylinder, gyroid to lamellar. The volume fraction of block B for a PA-b-PB diblock copolymer can be calculated like this:

$$\emptyset_B = \frac{N_B}{N_A + N_B} \qquad (2\text{-}8)$$

Where $N_A$ and $N_B$ is the degree of polymerization of polymer block A and B, and the volume fraction of block A can be calculated according to $\emptyset_A = 1 - \emptyset_B$.

Generally, when volume fraction $f_B < 0.25$, it forms sphere structure with A block as the matrix. When $f_B = 0.5$, it forms lamellar structure with both A and B block are equal. When $f_B \approx 0.3$, it forms cylinder structure with hexagonal, cubic or rectangular packed pattern. Rarely when $f_B = 0.36 \sim 0.39$, in this case it can form gyroid structure [25].

## 2.1.4 Thermoresponsive diblock copolymers

The thermoresponsive diblock copolymers is kind of stimuli-responsive diblock copolymers, which changes itself the structure and properties when the system is stimulated by external physical condition like temperature, pH, pressure and so on. In this thesis the focus is on the investigation of thermoresponsive diblock copolymers. Besides the general properties and structure of diblock copolymer, the thermoresponsive block can swell or de-swell according to the temperature changing above the lower critical solution temperature (LCST) or below LCST. The representative thermoresponsive diblock copolymer is poly(N-isopropylacrylamide) (PNIPAM), which has a LCST at 32°C in aqueous solution [16]. When heated in water above 32 °C (90 °F), it undergoes a reversible lower critical solution temperature (LCST) phase transition from a swollen hydrated state to a shrunken dehydrated state, losing about 90% of its volume [16]. When the PNIPAM is linked with PS as a PS-b-PNIPAM diblock copolymer, it still keeps this thermoresponsive property while PS keeps rigid regarding the temperature changing.

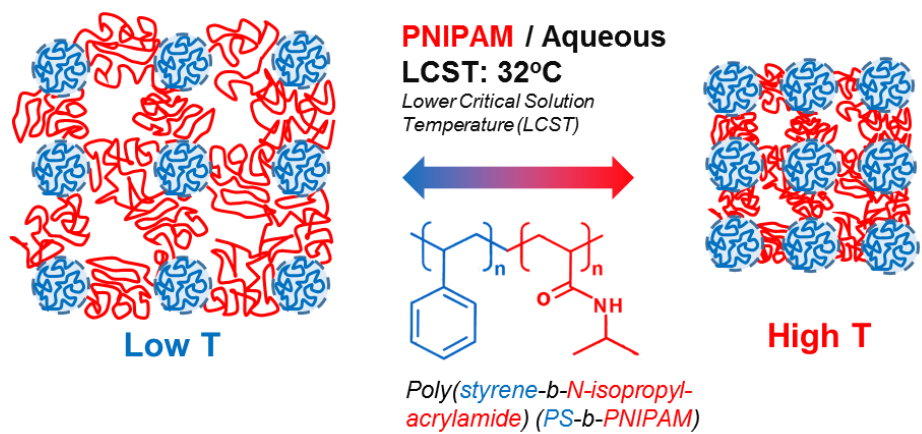

Figure 2-6: Illustration of thermoresponsive polystyrene block poly(N-isopropylacrylamide) (PS-b-PNIPAM) diblock copolymer swelling and deswelling state at low and high temperature respectively. [26]

## 2.2 Fundamentals of Magnetism

### 2.2.1 Basic concepts

The magnetism is a physical phenomenon, which describes the response of materials in magnetic field. According to different parameters like temperature, pressure and etc., materials can have different behaviors of magnetism. The process that a non-magnetic material obtaining magnetism is called magnetization, otherwise is called demagnetization [27]. A magnet can generate magnetic field due to the self-spin of unpaired electrons inside. When these magnetic fields have the same orientation, the material behaves as a magnet.

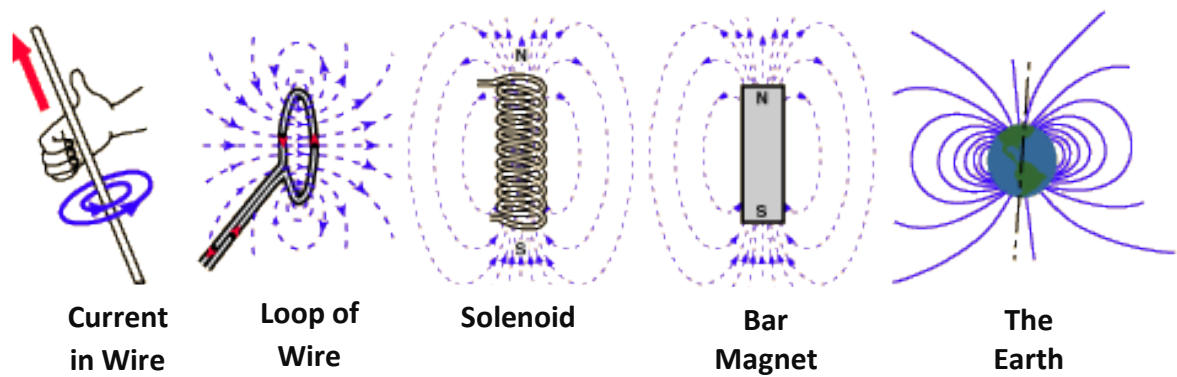

**Current in Wire**    **Loop of Wire**    **Solenoid**    **Bar Magnet**    **The Earth**

Figure 2-7: Sources of magnetic field: current in wire, loop of wire, solenoid, bar magnet and the earth. [28]

A magnetic field is the magnetic effect of electric currents and magnetic materials. It is a vector field with both direction and strength at any position in the space. There are two different vector field are used to describe the magnetic field, which are called magnetic field intensity (denoted as $H$) and magnetic flux density (denoted as $B$). The definition of magnetic field intensity H is: [29]

$$H \stackrel{\text{def}}{=} \frac{B}{\mu_0} - M \tag{2-9}$$

Where $\mu_0$ is magnetic constant, and M is magnetization. For linear materials the magnetization $M$ is proportional to $B$, thus $H=B/\mu'$, where $\mu'$ is material dependent parameter called the permeability.

## 2.2.2 Classification of Magnetic Materials

Magnetic materials can be classified according to the magnetic susceptibility $\chi_m$ values. In general, the types of magnetic materials are: [30]

(1) $-10^{-6} < \chi_m < -10^{-5}$. The materials' susceptibility is negative, and it is considered as diamagnetic material. Most diamagnetic materials have a very small negative susceptibility while superconductors are a unique case. In superconductors, due to the superconducting effect $\chi_m = 1$, thus it is very useful for applications like magnetic levitation.

(2) $10^{-5} < \chi_m < 10^{-3}$. The materials' susceptibility is small but positive, and it is considered as paramagnetic materials, such as aluminum, chromium, sodium, titanium, zinc, liquid oxygen, uranium, platinum, calcium, barium and so on.

(3) $\chi_m > 1$. The materials' susceptibility is usually quite large and positive, and it is considered as ferromagnetic materials, for example: Fe, Co, Ni, Gd, Dy, MnAs, $CrO_2$, $Fe_3O_4$, $NiFe_2O_4$. The ferromagnetic materials have a critical temperature called Curie temperature, where the spins align paramagnetically to an external field.

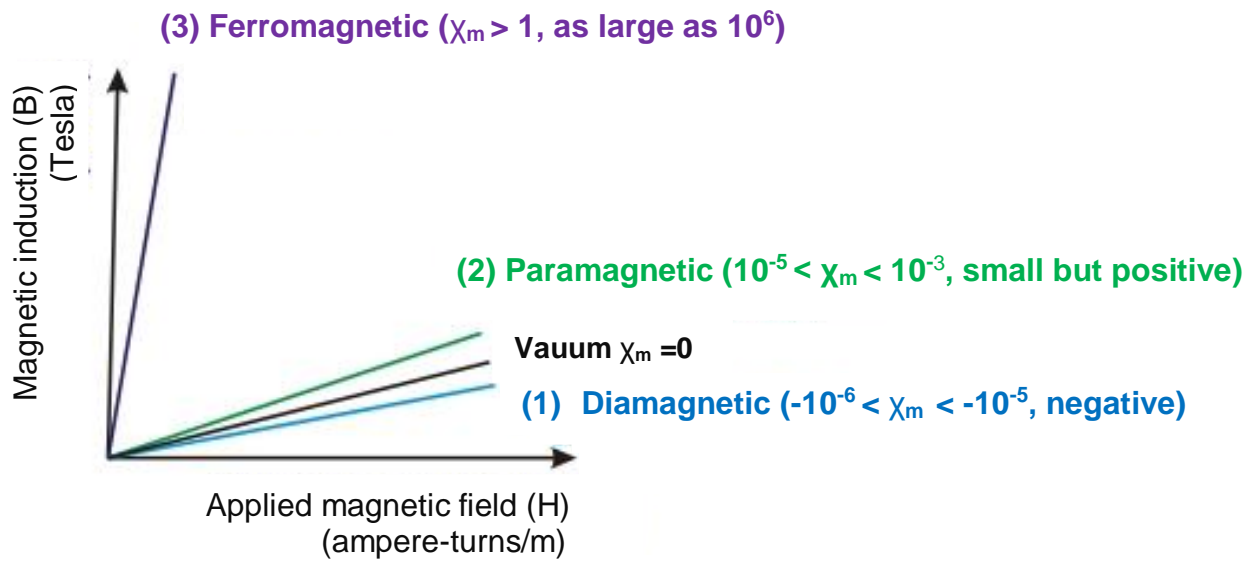

Figure 2-8: Classification of materials on the basis of susceptibility [30].

# 2.3 X-ray Scattering Techniques

## 2.3.1 X-rays and matters

X-rays are defined as electro-magnetic waves with much shorter wavelength (~0.1 nm) than that of visible light (~500 nm), which makes it possible to probe structures much smaller than what can be observed in a microscope. X-rays with high photon energies (above 5–10 keV, below 0.2–0.1 nm wavelength) are called hard X-rays, while those with lower energy are called soft X-rays. The hard X-rays are usually employed for materials imaging while the soft X-rays are easily absorbed by air or water with very short attenuation length [31].

The interaction of X-rays with matters are quite different compared with other radiological wave like ultraviolet, visible, infrared and microwaves. There are mainly two interactions between X-rays and matters: absorption and scattering, including Compton and Ryleigh scattering [31].

## 2.3.2 X-rays and structures

When X-rays are shined on the atoms, spherical waves comes out from the scattered position. Depending on the obersrevation angle 2θ , the orientation and the distance r, the interenfeces can be constructive meaning in phase, destrctive meaning out of phase. [32]

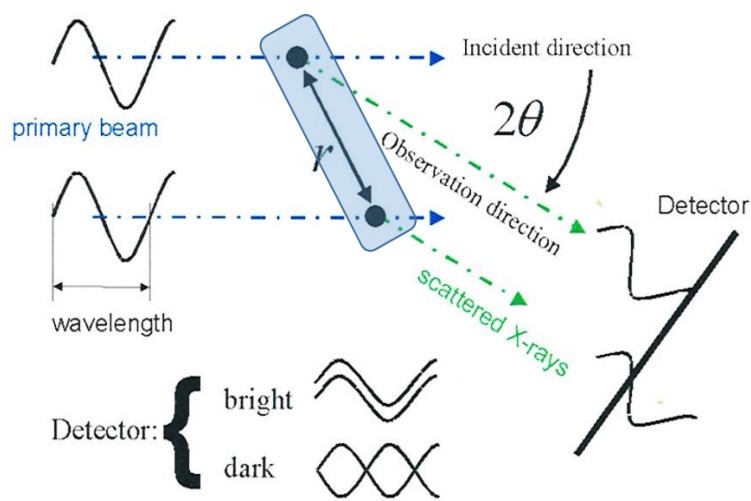

Figure 2-9: Illustration of X-ray interactions with structure and the detector brightness observation. [32]

The constructive and destructive interferences give different scattering patterns on the detector: bright spot and dark spot respectively. The scattering patterns can be expressed as a function of scattering angle $\theta$ and wavelength $\lambda$: [32]

$$q = \frac{4\pi}{\lambda} \sin(\theta) \qquad (2\text{-}10)$$

Where the $q$ is the momentum transfer or length of the scattering vector with the dimension is on over length.

### 2.3.3 Form factor and structure factor

The form factor is the measurement of the scattering amplitude of X-rays by a particle consists of many atoms. It is calculated by pair-distance distribution function $p(r)$: [32]

$$P(q) = 4\pi \int_0^\infty p(r) \frac{sin(qr)}{qr} dr \qquad (2\text{-}11)$$

Where $r$ is the distance between particles. The observed scattering pattern corresponds to the form factor only when the particles are far away from each other or identical in shape. When the sample is dilute, the scattering pattern is the form factor multiplied by the number of particles which are illuminated by the X-ray beam. If the particles have various sizes, the form factor is calculated by summing up to achieve the scattering pattern for the sample [32]. The structure factor is an additional interference pattern multiplying the form factor of the single particles. Especially in crystallography it is called lattice factor which includes all the information regarding the positions of particles with respect to each other. At small angle scattering, the concentration effects become visible [32]. When the particles are highly ordered, there is a pronounced peak which is called a Bragg peak. The Bragg's law gives the distance between ordered particles in the following formula. [33]

$$d_{Bragg} = \frac{2\pi}{q_{Bragg}} \qquad (2\text{-}12)$$

# Chapter 3

# Sample Preparation

## 3.1 Materials

### 3.1.1 Thermoresponsive diblock copolymer

To fabricate desired metal-polymer hybrid nanocomposites, the thermoresponsive polystyrene-*block*-poly(N-isopropyl acrylamide) (PS-*b*-PNIPAM) diblock copolymers (DBCs) with two different Number Average Molar Mass ($M_n$) and PNIPAM volume fraction were used, which are purchased from *Polymer Source Inc*. The polymer details are presented in the following table 3-1 and the chemical structure of PS-b-PNIPAM DBCs is as shown in figure 3-1.

| Polymer | $M_n$ [kg/mol] | Mw/Mn (PDI) | $f_B$ | $T_g$ [$^oC$] | Sample # |
|---------|----------------|-------------|-------|--------------|----------|
| <u>PS</u>-*b*-PNIPAM | 16.0-b-9.5 | 1.3 | 0.42 | 107-181 | P14514B-SNIPAM |
| PS-*b*-<u>PNIPAM</u> | 11.5-b-24 | 1.2 | 0.68 | | P14966-SNIPAM |

Table 3-1: Properties of two kinds of PS-*b*-PNIPAM DBC: number average molecular mass $M_n$, polydispersity index (*PDI*), volume ratio of block PNIPAM $f_B$, glass transition temperature $T_g$ [34-35], and sample series number from *Polymer Source Inc*.

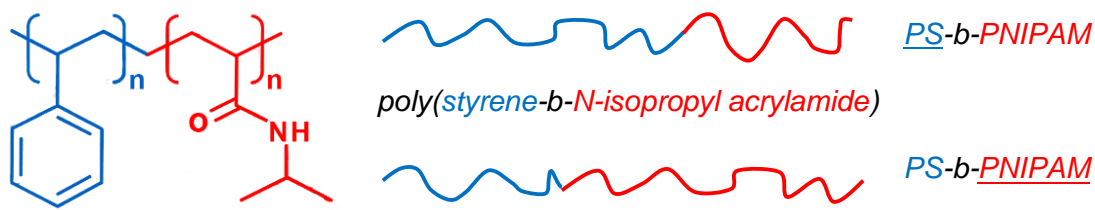

Figure 3-1: The chemical structure of investigated two kinds of PS-*b*-PNIPAM DBCs.

## 3.1.2 Iron salt

The iron oxide/DBC hybrid samples can be prepared either by annealing iron salt containing DBC or by incorporating PS-coated iron oxide nanoparticles (NPs) into the DBC. The organic iron salt can be dissolved in methanol and mixed with the DBC solution to distribute the iron salt evenly inside the polymer matrix. In this thesis, the used organic iron salt is synthesized *iron (II) chloride (2, 2'- dipyridyl) complex [FeCl$_2$(2, 2'-dipyridyl)]*. The chemical structure is shown in figure 3-2.

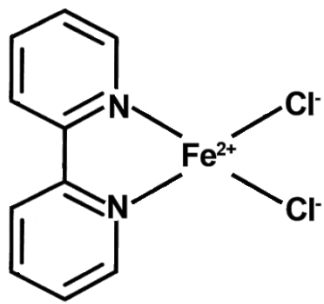

Figure 3-2: The chemical structure of *FeCl$_2$(2, 2'-dipyridyl)*

Due to the polarity of the iron salt, it accommodates within the PNIPAM block of PS-b-PNIPAM DBC and finally becomes iron oxide embedded inside PNIPAM block after heat post-treatment.

### 3.1.3 Iron oxide nanoparticles

The iron oxide NPs with surface polymer modification can be embedded in the DBC matrix. The iron oxide nanoparticles used in this thesis are $Fe_2O_3$ nanoparticles modified with PS-chains on the surface. The $Fe_2O_3$ nanoparticles are synthesized and dissolved in toluene with a density of 1.0 g/cm$^3$ and size around 10 nm. Thus it goes to the PS block in the PS-*b*-PNIPAM block structure according to the solubility principle.

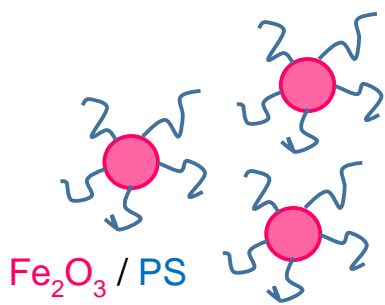

Figure 3-3:  A schematic view of PS surface modified $Fe_2O_3$ nanoparticles.

### 3.1.4 Substrates

The substrates used in this thesis is silicon substrates and mica window. The silicon substrate usually undergoes either acid cleaning or basic cleaning to remove the surface impurities and oxidized silica and established hydrophobic or hydrophilic surface for films.

**a) Acid cleaning protocol**

The aim of acid cleaning is to remove the silica and impurities on the surface during the cutting of silicon wafer. A beaker is put in a water bath heating until 80ºC, and 54 mL deionized $H_2O$ and 84 mL $H_2O_2$ (30%) are added into the beaker and mixed during the heating. Then 198 ml concentrated $H_2SO_4$ (96%) is added into the beaker slowly by stirring. Ensure the liquid level inside and outside the beaker is similar. Waiting until the whole system is heated upon 80ºC, the pre-cut silicon wafers mounted on a Teflon sample holder can be immersed into the acid cleaning solution for 15 min. After taking out the Teflon sample holder with silicon substrates, rinse them with deionized water (DI) for three times and dry them with nitrogen blow (pressure ~ 2 bar).

**b) Basic cleaning protocol**

The aim of basic cleaning is to remove the silica and impurities on the surface during the cutting of silicon wafers and established a hydrophilic 1 nm silicon oxide layer at the silicon surface [13]. The pre-cut silicon substrates are immersed in dichloromethane (purity ≥99.8%) at 46ºC for 30 min, then they are rinsed by DI water. To remove the impurities and organic traces, Si substrates are mounted on a Teflon sample holder and placed into a basic bath. The beaker is filled with 30 mL $NH_3$ (30%), 30 mL $H_2O_2$ (30%) and 350 mL DI water, then the basic bath is heated upon 76ºC and kept for 2 hours. Afterwards, take out the Teflon sample holder with Si substrates and rinse them with DI water, followed by drying them with nitrogen blow (pressure ~ 2 bar) [13].

# 3.2 Nanocomposite preparation

## 3.2.1 Solution preparation

- **PS-b-PNIPAM DBC / iron salt solution**

The solution for PS-b-PNIPAM DBC / iron salt nanocomposites is prepared by using tetrahydrofuran (THF, purity ≥99.8%) as solvent for PS-b-PNIPAM DBC and methanol (purity ≥99.5%) as solvent for iron salt. A fixed polymer concentration of 80 mg/mL is used for solution casting method and 40 mg/mL for spin coating method. The solutions of PS-b-PNIPAM DBC / iron salt with different contents of iron salt are prepared according to the molecular ratio of Fe salt and NIPAM molecules (0, 0.01, 0.03, 0.05, 0.1. 0.3, 0.5). And the total volume is fixed as 1 mL with 0.9 mL THF and 0.1 mL methanol for all solutions with different [Fe]/[NIPAM] ratio.

- **PS-b-PNIPAM DBC / $Fe_2O_3$ nanoparticles solution**

The solution for PS-b-PNIPAM DBC / $Fe_2O_3$ nanoparticles is prepared by using tetrahydrofuran (THF, purity ≥99.8%) as solvent for PS-b-PNIPAM DBC and toluene (purity ≥99.5%) as solvent for $Fe_2O_3$ nanoparticles. The same fixed polymer concentration of 80 mg/mL is used for solution casting method and 40 mg/mL for spin coating method. The solutions of PS-b-PNIPAM DBC / iron salt with different content of $Fe_2O_3$ NPs are prepared according to the weight ratio (wt%) of Fe$_2$O$_3$ NPs and PS block (0, 0.05, 0.1, 0.3, 0.5. 1, 3, 5). And the total volume is fixed as 1 mL with 0.9 mL THF and 0.1 mL toluene for all solutions with different Fe$_2$O$_3$ NPs/PS ratio.

## 3.2.2 Solution casting method for bulk sample preparation

To fabricate bulk nanocomposite film, we employ the solution casting method which is an easy and controllable method for preparation of bulk sample. There are mainly three steps: solution dropping, solution casting and thermal annealing as shown in figure 3-4.

1) <u>Solution Dropping:</u> The prepared PS-b-PNIPAM DBC dissolved in tetrahydrofuran solution and the *FeCl$_2$(2,2'-dipyridyl)* dissolved in methanol or *Fe$_2$O$_3$* nanoparticles dissolved in toluene are mixed together with different ratio and wobbled for one hour.

2) <u>Solution Casting:</u> A mica window with diameter of 6 cm is mounted in a stainless steel cell which has a hole with diameter of 3 cm for X-ray passing through. Then a rubber o-ring is put on the mounted mica window and the prepared solution is dropped on the mica window using a pipette. Each time the volume of dropped solution is 20 μL and the next drop is after 20 min to make sure the solution is completely dry until the prepared solution is used up. The remaining solvent in the bulk film may cause a lot of bubbles after thermal annealing. These preparation steps were performed at room temperature in ambient conditions.

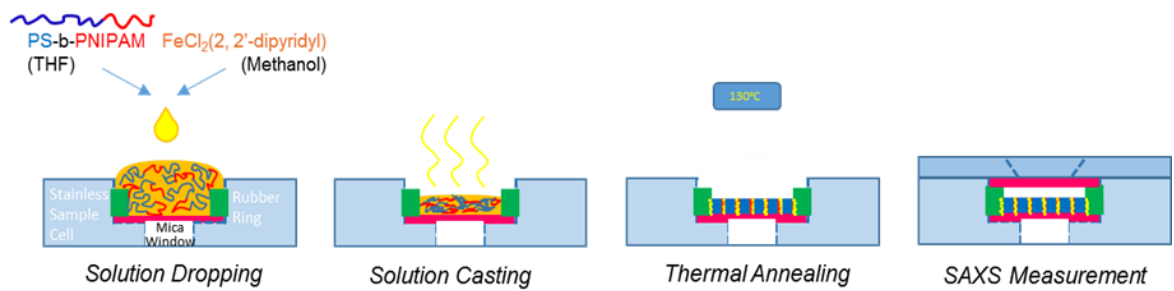

*Solution Dropping*     *Solution Casting*     *Thermal Annealing*     *SAXS Measurement*

Figure 3-4: Schematic diagram of solution casting method where bulk samples are sandwiched between two mica windows.

3) <u>Thermal Annealing:</u> The prepared cell with bulk inside can be shifted to a constant temperature drying oven and annealed for 48 hours at 130 ºC in a nitrogen atmosphere. Finally, quickly take out the cell and cool it down to room temperature, then put another mica window on top and close the cell with six screws. The cell is ready for in-house SAXS measurement now. And if not use it immediately, the sample should be stored in glove box after preparation in case of ambient humidity influence.

### 3.2.3 Spin-coating method for thin film preparation

To fabricate metal oxide/DBC hybrid thin film, we employ the spin-coating method which is a widely used technique for preparing evenly thin films on silicon or glass substrates. There are main three steps: solvent cleaning, spin coating and thermal annealing as shown in figure 3-5.

1) <u>Solvent Cleaning:</u> The pre-cleaned silicon substrate needs to be further rinsed with solvent prior to coating step. The drop of solvent should cover all the area of the silicon substrate. The setting parameters for spin-coater are 2000 rpm, spin time 30s and acceleration in 9s.

2) <u>Spin Coating:</u> After solvent spin-dry, the prepared mixed solution is dropped on the silicon and make sure it covers all the area of silicon substrate. The same above spin coating parameters were employed.

3) <u>Thermal Annealing:</u> The spin-coated thin films were moved in an oven and annealed them for 48 hours at 130 ºC in a nitrogen atmosphere. The sample should be stored in glove box after preparation in case of ambient humidity influence.

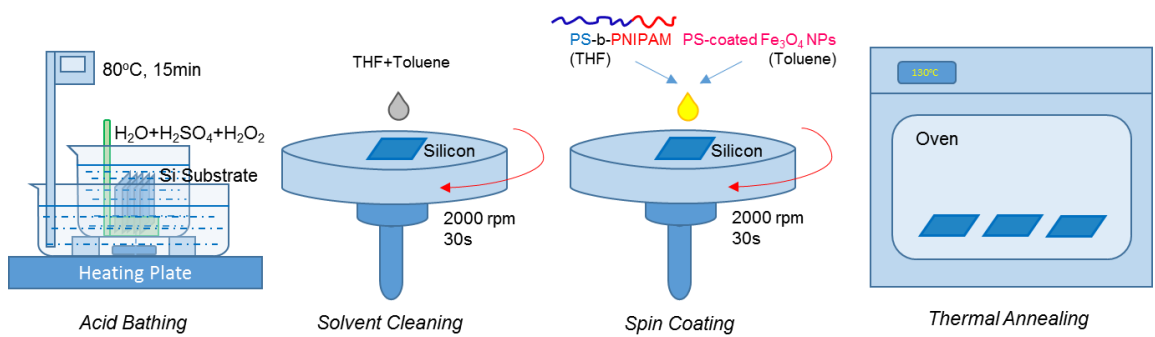

Figure 3-5: Schematic diagram of spin-coating method.

# Chapter 4

# Characterization Methods

In this thesis, several characterization methods are employed for different aims on the investigation. First the mesoscopic block morphology of the PS-b-PNIPAM DBC thin film are probed by scanning electron microscope (SEM) due to its high resolution on the surface. Then for the structural characterization of bulk nanocomposites, the small angle X-ray scattering (SAXS) technique is used while for the thin film nanocomposites, the grazing incidence small angle X-ray scattering (GISAXS) technique is employed. Finally, the superconducting quantum interference device (SQUID) is used for the investigation of magnetic properties of metal oxide/DBCs nanocomposites.

## 4.1   Morphology Characterization

### 4.1.1 Scanning electron microscope (SEM)

The scanning electron microscope (SEM) uses a focused beam of high-energy electrons to scan the surface of solid sample, then it collects the signals of secondary electrons (SE) and back scattered electrons (BSE), and transfer them info SEM image. A scanning electron microscope consists of three parts: vacuum system, focused electron beam system and imaging system.

The basic principle is that the electron gun generates electron beam which passes through anode then focused by a magnetic lens on the solid sample surface. The high energy electron beam hits the sample and brings secondary electrons, back scattered electrons, Auger electrons, X-rays and so on, thus an imaging system including different detectors for different signals needs for SEM imaging.

In general, the SEM can only analyze the sample surface due to the penetration length of the electron beam is very small. The secondary electrons give the information of the nanoscale surface roughness and the back scattered electrons give the information about the order of atoms on the surface. The SEM measurements were performed with a NVision40 FESEM by Carl Zeiss AG and the accelerating voltage is 2 kV with a working distance of 1.2 mm. The Gwyddion 2.37 software was used for image processing.

## 4.2   Structural Characterization

### 4.2.1 Small angle X-ray scattering (SAXS)

The small angle X-ray scattering (SAXS) is a small angle scattering (SAS) technique where the X-rays are scattered inhomogeneously in the nanometer scale at very small angle. Typically, the scattering angle for SAXS is 0.1 ~ 5°. The shape, size and structure of macromolecules can be probed by the scattered X-rays in this angle range. Generally, the SAXS technique can probe the structural information of macromolecules with a dimension between 5 and 25 nm. Thus it is a quite powerful technique for the investigation of soft matters like polymers. Depending on the distance between sample and detector, if it is very near that means the scattering angle is quite large, this technique is called wide angle X-ray scattering (WAXS) [36].

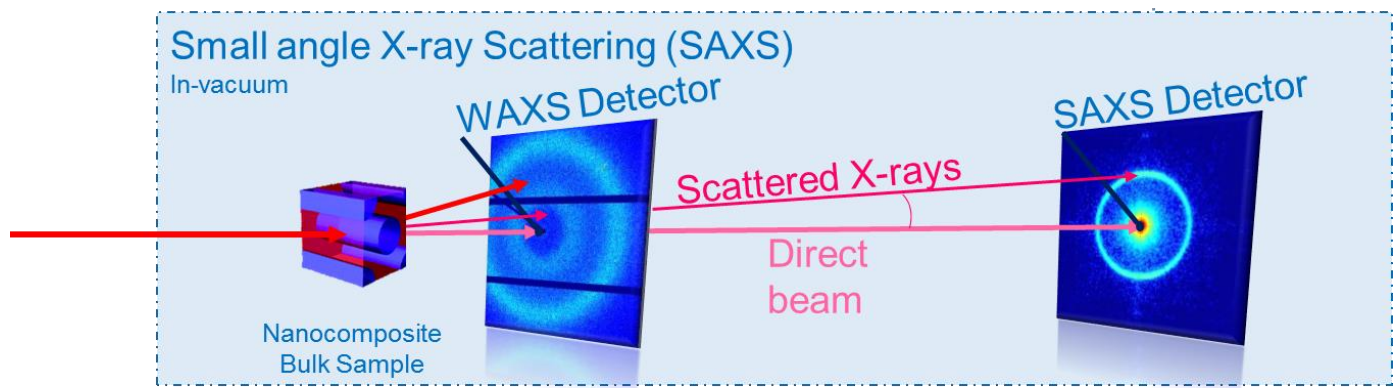

Figure 4-1: Geometry of small angle X-ray scattering technique.

### 4.2.1.1 Ganesha SAXS instrument at TUM

A SAXS instrument consists of an X-ray source, a collimation system, a sample, a beam-stop and a detection system. The collimation system is used for narrowing the beam and defining the zero-angle positon. And the beam-stop is positioned before the detector to stop the direct beam and protect the detector. Besides, by covering the direct beam, it also improves the relatively weak scattering signal from the sample and makes the contrast better. For an in-house sample instrument, the X-ray path and the sample should be in vacuum to decrease the background scattering by the air. In this thesis, the SAXS experiments are performed with a Ganesha SAXS instrument in the SAXS lab of Chair of Functional Materials at Technical University of Munich.

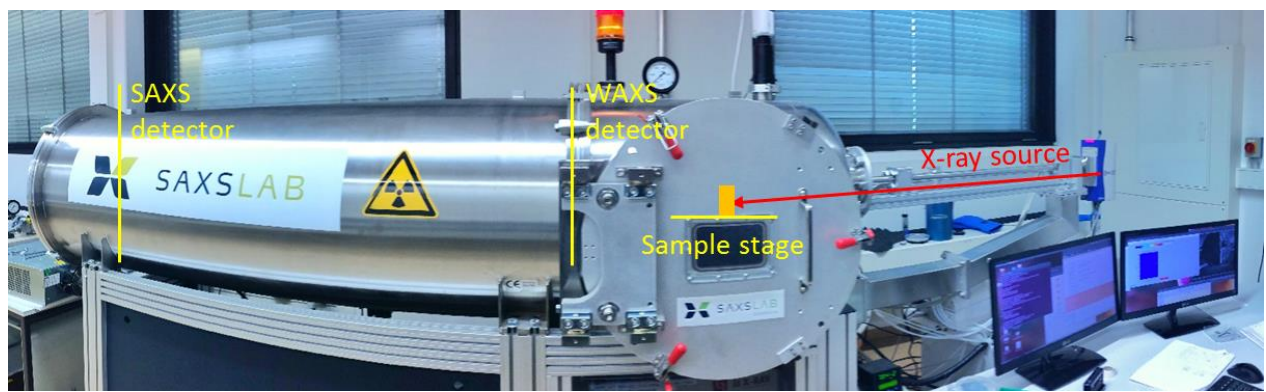

Figure 4-2: The Ganesha SAXS instrument in SAXS lab of Chair of Functional Materials at TUM.

The Ganesha SAXS instrument consists of X-ray source (49.8 keV, 0.59 mA) and controller (Genix$^{3D}$), motor drived sample stage, flying chamber (1.5 m), beam-stops, Pilatus 300k detector (pixel size 172 μm × 172 μm), vacuum pump and software controlling system. The sample to detector distance is 1050 mm and the radiation wavelength λ = 0.154 nm. To start the experiment, the chamber should be vented first and then the sample should be placed correctly on the stage. Move the sample stage to check if it can be freely positioned using the commands in the SAXS terminal and Graphical User Interface (GUI). Then evacuate the chamber until the vacuum reaches $10^{-2}$ mbar, move the sample stage out of the beam path and configure the sample stage distance, the SAXS beam-stop, the detector distance. Now move beam-stop to the direct beam position and then move the stage back to the beam path to check the sample center. After everything done correctly, the SAXS instrument is ready for SAXS measurement.

### 4.2.1.2 Temperature and humidity controlling cell

For precisely controlling of the temperature and humidity conditions of the thermoresponsive system, a temperature and humidity controlling cell was designed and constructed.

- **Mechanical design**

The mechanical design for the temperature and humidity controlling cell considered sample holder, X-ray beam path, scattering angle limitation, vacuum sealing, water cycling and humidity space.

1) Sample Holder: The bulk sample is prepared as a free standing film which is mounted in the hole with diameter of 4 mm of an aluminum thick foil. Then the sample holder is vertically placed in the center of the cell with two grooves on each side of the cell wall.

2) X-ray Beam Path: The X-ray goes through the center hole of the mounted sample holder in the cell, thus two Kapton windows (polyimide film) are employed in the front and back sides due to its high mechanical and thermal stability and high transmittance to X-rays. And two aluminum frameworks with four screws are used to fix the Kapton windows on the cell.

3) Scattering Angle Limitation: The size of the Kapton window and the distance between sample holder and Kapton window confines the scattering angle of the scattered X-ray. In out designed cell, the distance from sample holder to window is 1.9 cm and the width of the Kapton window is 1.5 cm. Thus the calculation for the maximum scattering angle is:

$\theta$ (in degree) = arctan (0.75 cm/1.9 cm) = arctan (0.3947) = 21.54° > 5° (MAXS) >> 2° (SAXS)

According to the calculated result, the scattering angle is much larger than the requirement for middle or small angle scattering. While for wide angle scattering which requires for scatter angle from 5° to 60°, the sample holder needs to be mounted closer to the Kapton window on the detector side.

4) <u>Vacuum Sealing:</u> The vacuum degree issue is vital for the scattering background statistically. The junctions of the cell are sealed with rubber rings instead of using vacuum glue. And the vacuum in the Ganesha in-house SAXS instrument can reach to $10^{-2}$ mbar which is enough for performing small angle X-ray scattering experiments.

5) <u>Water Cycling:</u> The cell has feedthrough drills in the aluminum cap part and body and it is isolated from the inner humidity space. On one side they are connected by Festo push-in fittings and tubes, on the other side they are connected to the Julabo thermostat system.

6) <u>Humidity Space:</u> The space inside the cell are connected with the humidifier through Festo push-in fittings and tubes. There is also a small volume reservoir in the bottom of the inner space, which is for keeping the initial humidity circumstance inside the cell.

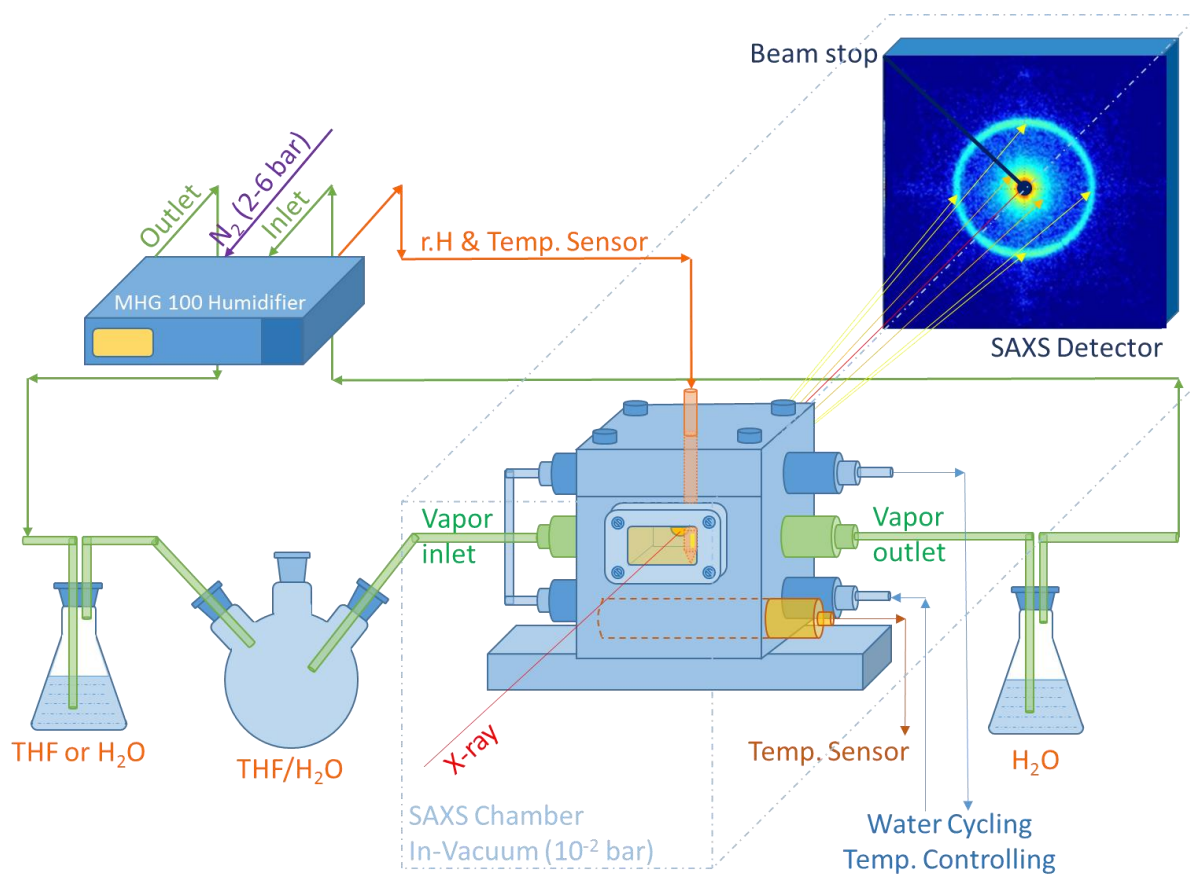

Figure 4-3: Schematic design of the humidity and temperature controlling cell.

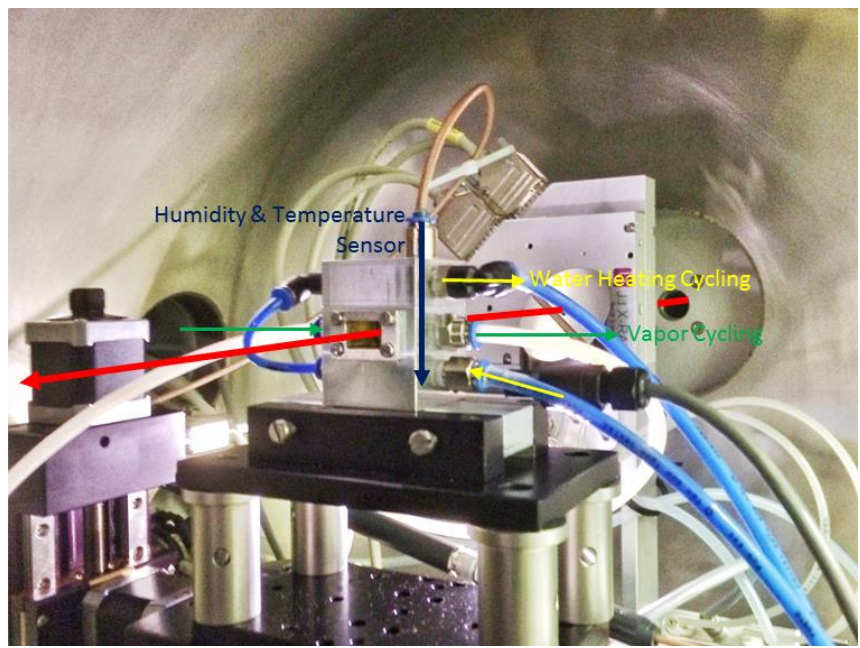

Figure 4-4: Assembled humidity and temperature controlling cell in Ganesha SAXS instrument.

- **Electronic design**

Since the temperature and humidity controlling cell will be within the Ganesha in-house SAXS instrument chamber under vacuum circumstance, the sensor electronic connection needs to be re-designed to go through the electronic feedthrough on the back side of the SAXS instrument.

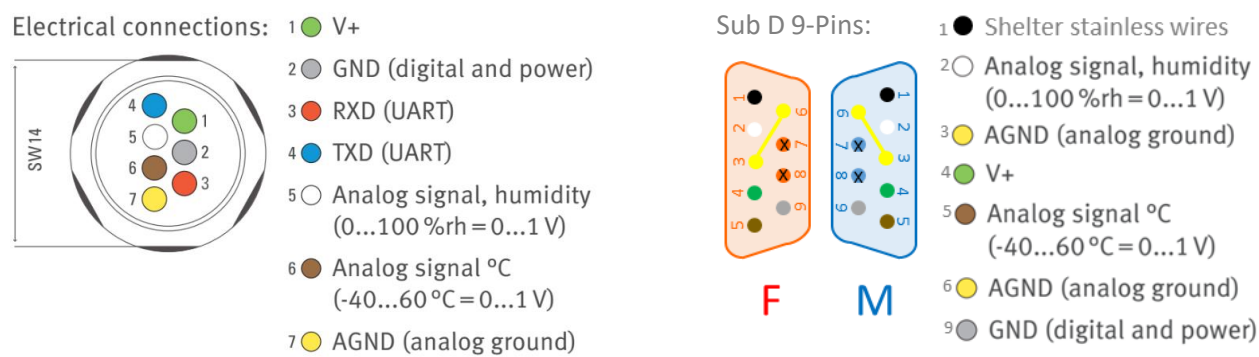

Figure 4-5: Electronic pin configurations of the sensor connection cable.

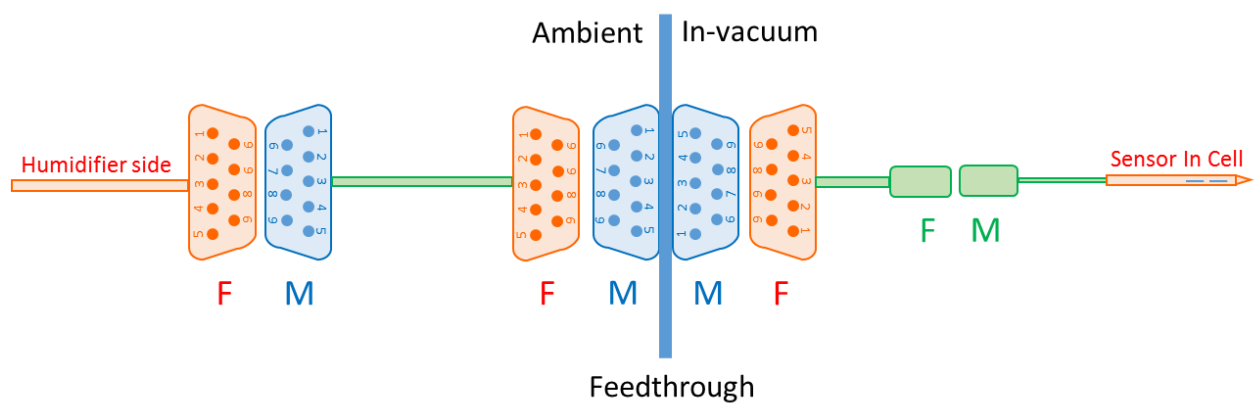

Figure 4-6: Schematic of the electronic connections of sensor cable.

- **Temperature regulation**

The temperature regulation is controlled by the Julabo thermostat system. It provides precisely controlling of water temperature and regulate the temperature for the system according to the Julabo temperature sensor. The temperature and heating or cooling rate can be set in the command window of Ganesha in-house SAXS instrument.

- **Humidity regulation**

The humidity can be either automatically regulated by the MHG 100 humidifier or manually controlled by injecting water to the reservoir of the cell.

1) <u>MHG 100 Humidifier Regulation:</u> This humidifier can generate vapor and regulate the relative humidity in the cell according to the connected humidity and temperature sensor. The absolute humidity is measured and the relative humidity is calibrated depending on the real temperature acquired from the sensor. The humidity curve can be generated from a software called MHG Method Editor, then the humidifier can be connected to the laptop using a cross-linked RS232-USB cable and programmed according to the humidity curve in the software called MHG Control.

2) <u>Water Injection Controlling:</u> Due to the limited regulation ability of the humidifier at high temperature, the water injection controlling method is also used when increasing the temperature of the cell. The water reservoir is just below the end of the vapor in-let tube. Thus an injector is used for adding sufficient amount of water in the reservoir to make sure the sample is in the humidity circumstance. And nitrogen flow is used for reducing the vapor inside the cell. Thus we controlled the humidity inside the cell by adding certain amount of water. The real humidity value is still collected by the sensor connected to the humidifier.

### 4.2.1.3 Ratio of characteristic peak positions for ordered structures in SAXS 1D profiles

The 2D SAXS pattern can be achieved and open in the SAXS GUI software, and there is also automatically processing function to process the 2D SAXS pattern by averaging the intensity of scattering ring and plot the intensity versus $q$ value in a 1D plot. The ratio of scattering peak relative positions on q-scale gives the information of the possible structure.

| Ordered Structure | Relative q-scale  peak position ratio |
|---|---|
| Cubic | $1 : \sqrt{2} : \sqrt{3} : 2 : \sqrt{5} \dots$ |
| Lamellar | $1 : 2 : 3 : 4 : 5 \dots$ |
| Hexagonal | $1 : \sqrt{3} : 2 : \sqrt{7} : 3 \dots$ |

Table 4-1: Relative q-scale peak position ratio for cubic, lamellar and hexagonal ordered structures.

## 4.2.2 Grazing incidence small angle X-ray scattering (GISAXS)

The grazing incidence small angle X-ray scattering is also a SAS technique, which is similar to SAXS but with grazing incidence mode. This means the X-rays impinges on the sample surface at a very shallow angle below 1 degree. Thus it is considered as very powerful technique for the investigation of thin film samples. To enable the X-ray penetrates the film, the grazing incidence angle should be larger than the critical angle of the materials, otherwise only X-ray reflection exists.

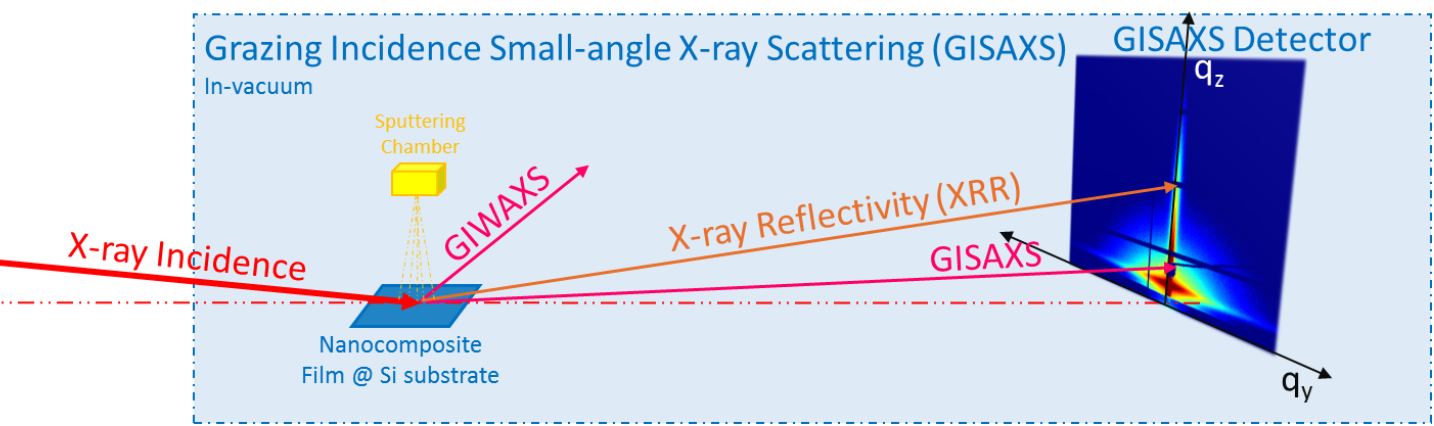

Figure 4-7: Geometry of grazing incidence small angle X-ray scattering technique.

In a GISAXS experiment, besides the same components used in the SAXS experiment, a special sample stage with two adjustable angles controlled by rotational motors is used to align the film sample. In addition, one beam-stop needs to stop the direct beam and another one needs to stop the specular beam which is quite strong due to the mirror reflection. Normally the in-house GISAXS experiment can take more than 12 hours to get enough statistic scattering event information on the detector because of the limited X-ray energy for in-house instrument. Therefore, the GISAXS experiments in this thesis are taken at the large scale synchrotron facility where sufficient X-rays fluxes are provided and the time for one high-quality pattern could be less than 0.1 second. This advantage brings the possibility to carry on in-situ GISAXS experiments like sputtering metal nanoparticles on the film surface and use GISAXS to probe the structure changes.

- **Static GISAXS experiment at Elettra**

The static GISAXS experiments were performed in the Australian SAXS beamline at Elettra Sincrotrone Trieste in Italy. Elettra is the third generation synchrotron with 2 and 2.4 GeV storage ring that has been in operation since 1993. The Australian GISAXS beamline is equipped with two detectors: a 1M Pilatus detector (984 x 1043 pixels) for GISAXS and a 100k Pilatus detector (487 x195 pixels) for GIWAXS with pixel size 172μm x 172 μm for both. And the beam-stops are positioned inside the flying tube, which is sealed with Kapton windows on both sides. The sample is positioned on a rotational stage between the beam shutter and the flying tube. The shutter is operated by the control panel in the controlling room, which is isolated by the safety gate made by concreate and lead. When the sample is positioned ready, the operator should get the safety key and close the gate within 30 seconds. For the static GISAXS experiments, the sample to detector distance is 1967.02 mm and the synchrotron radiation wavelength λ is 0.154 nm with 8keV photon energy. The beam size is narrowed as 0.5 mm x 0.2 mm. The incident angle is set as 0.43$^\circ$ and the exposure time is 300s.

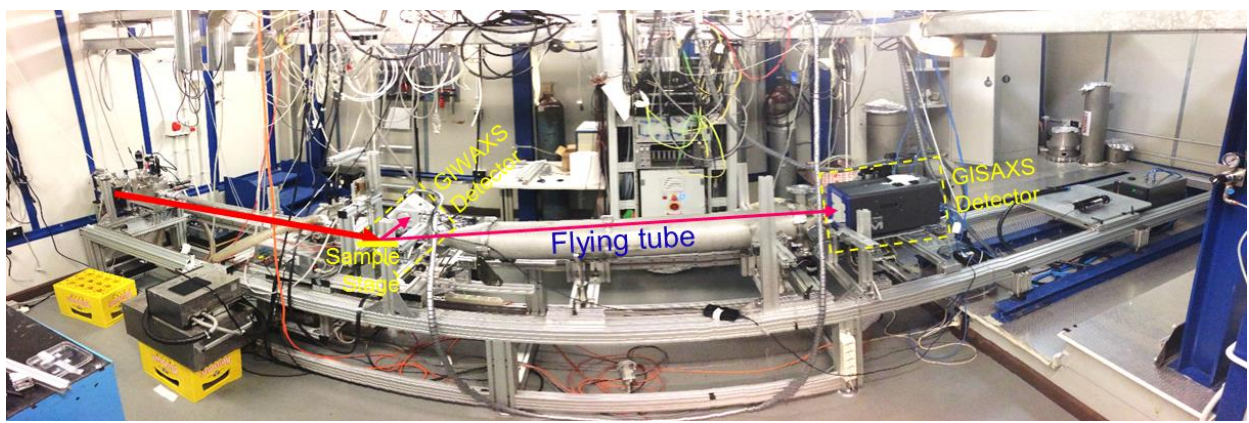

Figure 4-8: The Australian SAXS beamline at Elettra Sincrotrone Trieste.

- **Sputtering GISAXS experiment at DESY**

The sputtering GISAXS experiments were performed in P03 Micro- and Nanofocus X-ray Scattering (MINAXS) beamline of the PETRA III storage ring at Deutsches Elektronen-Synchrotron in Hamburg. It is equipped with a Pilatus 300k detector (487 ✕619 pixels, pixel size 172μm ✕ 172 μm) and an ultra-high vacuum sputter chamber which is mounted on a goniometer (Huber). The synchrotron radiation wavelength is 0.954 Å and the photon energy is 13 keV. The sample to detector distance equals to 3988 mm and the incident angle 0.4º is used to clearly separate the Yoneda peak of the involved materials (Au, Si, PS-*b*-PNIPAM) from the specular peak. The beam is narrowed to 38μm ✕ 19 μm by beryllium lenses. The sputter chamber deposits gold nanoparticles (99.999%, MaTeck GmbH) onto film surface with a sputtering rate at 0.007 nm/s. And it is operated at 150 w power with pre-evacuated vacuum at 0.02 mbar.

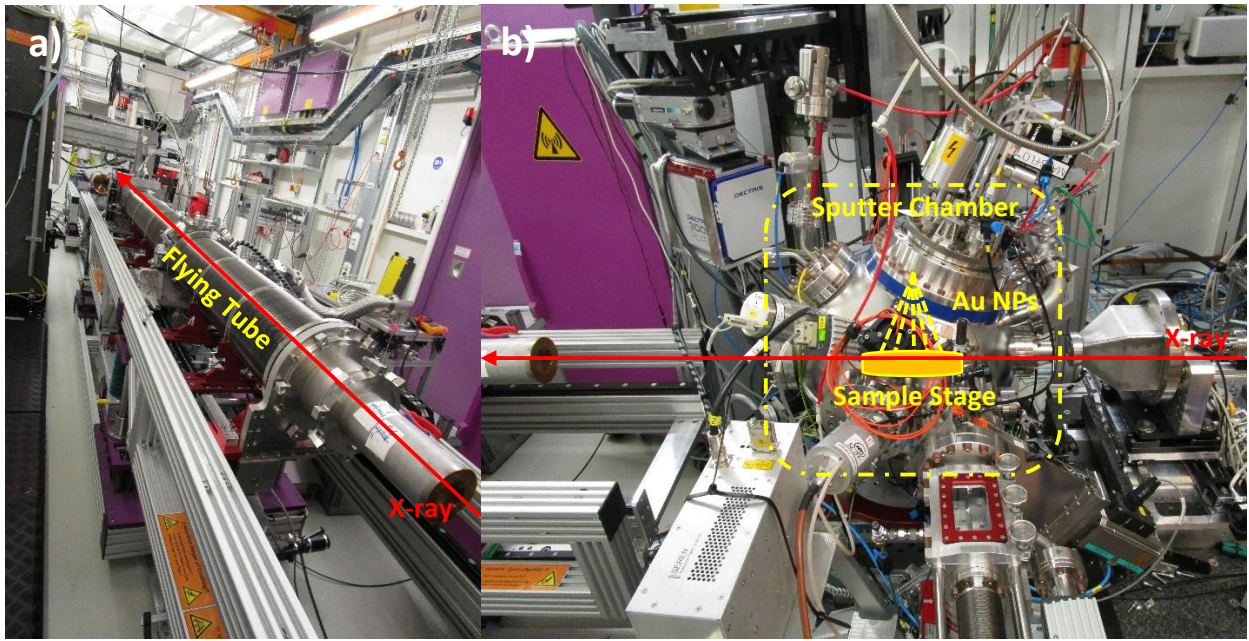

Figure 4-9: P03 MINAXS beamline of the PETRA III storage ring at DESY in Hamburg with a) axial view of flying tube and b) side view of sputter chamber.

## 4.3 Magnetic Properties Characterization

### 4.3.1 Superconducting quantum interference device (SQUID)

The Superconducting quantum interference device (SQUID) is a magnetometer which is very sensitive to measure extremely subtle magnetic fields as lows as 5 aT ($5 \times 10^{-18}$ T). Since the added iron salt or iron nanoparticles possess very weak magnetism, the SQUID magnetometer is capable to measure the magnetic behavior of the nanocomposites.

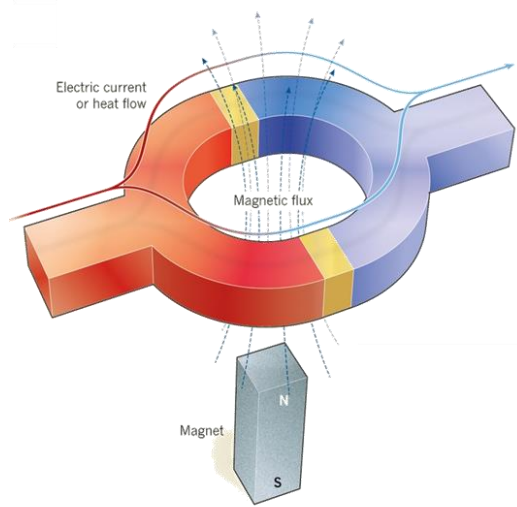

Figure 4-10: In DC -SQUIDs, a superconducting loop contains two Josephson junctions - thin insulating barriers (yellow) sandwiched between the two superconductors (red and blue) [37].

Mainly there are two kind of SQUID, direct current (DC) and radio frequency (RF). The RF SQUID magnetometer only needs a Josephson junction (superconducting tunnel junction) with lower productive costs compared with DC SQUID, but the sensitivity is also less than the DC SQUID. The DC type SQUID magnetometer has combined two parallel Josephson junctions into a superconducting ring, and the function is based on the Josephson Effect [37]. The magnetic property measurements were carried out with a MPMS XL-7 SQUID-Magnetometer (Quantum Design, San Diego, USA) in the lab of Bayerische Akademie der Wissenschaften. The magnetization is measured as a function of temperature. All measurements are carried out with an external magnetic field from -700 to 700 mT in the film plane and repeated at different temperatures of 2, 150, 300K.

# Chapter 5

# Characterizations of PS-*b*-PNIPAM DBC / Iron Oxide Hybrid Nanocomposites

In chapter 5, the investigation results of PS-*b*-PNIPAM DBC / iron oxide hybrid nanocomposites are presented. The PS-*b*-PNIPAM DBC / iron salt hybrid nanocomposites were investigated by static SAXS with a linkam heating stage. And the PS-*b*-PNIPAM DBC /iron oxide thin films were investigated by static GISAXS measurements at Elettra. The bare PS-*b*-PNIPAM thin film was investigated by sputtering GISAXS measurements at DESY. The first stage exploring experiments for thermoresponsive PS-b-PNIPAM DBC bulk and thin film were probed by using ex-situ and in-situ SAXS measurements with designed temperature and humidity controlling cell. The superparamagnetic properties were found for the PS-*b*-PNIPAM DBC / iron oxide hybrid nanocomposites by SQUID measurements.

# 5.1 Structural Characterizations of PS-b-PNIPAM DBC / Iron Oxide Hybrid Bulky Films: SAXS and SEM Study

### 5.1.1 Block structure

The block structures of the PS-b-PNIPAM DBC / iron oxide hybrid nanocomposites are investigated by small angle X-ray scattering technique (SAXS). A series of PS-b-PNIPAM DBC/iron oxide hybrid nanocomposites with different [Fe]/[NIPAM] molecular ratio from 0 to 0.5 are prepared according to the method described in chapter 3. The thermal post-treatment of bulk samples composed of iron salt and PS-b-PNIPAM DBC would result in not only forming nanostructured materials but also decomposing the iron salt into iron oxide. This has been previously approved for iron salt containing PS-b-PMMA DBC [38]. The SAXS experiments were performed with the Ganesha in-house SAXS instrument. Each SAXS profile was acquired for one hour.

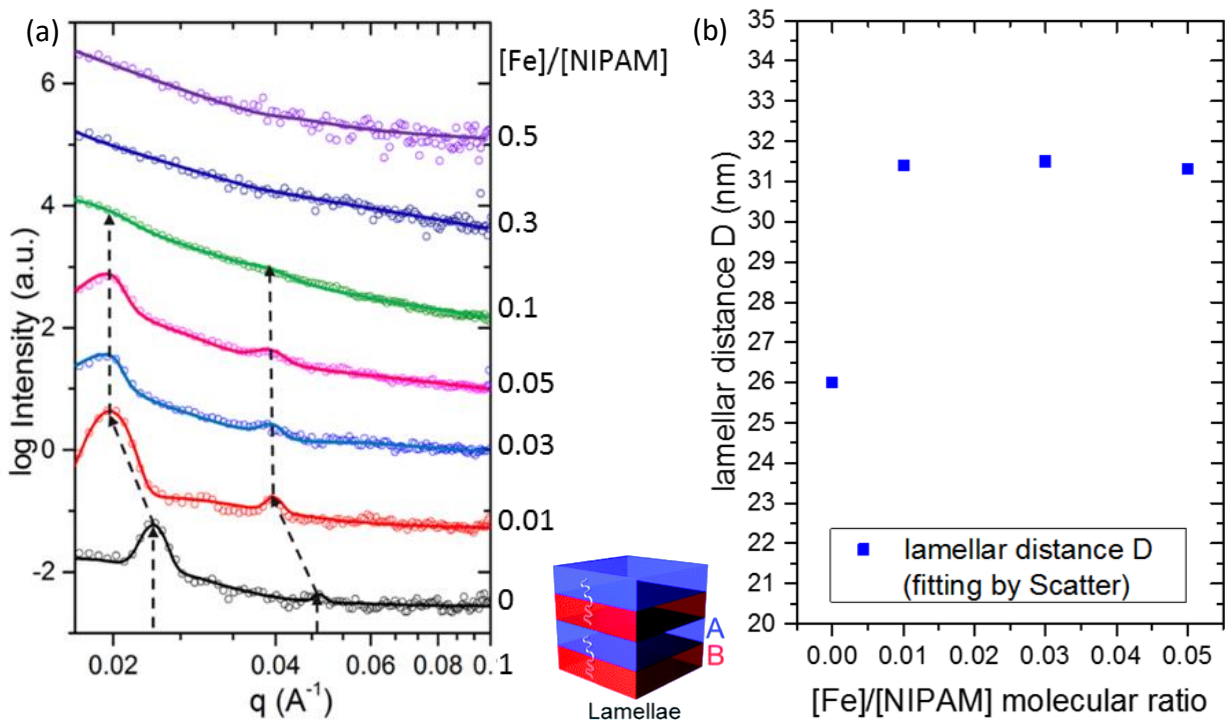

Figure 5-1: (a) SAXS profile of iron salt/ PS-b-PNIPAM nanocomposites with [Fe]/[NIPAM] molecular ratio increasing from 0 to 0.5. (b) Lamellar distance D versus [Fe]/[NIPAM] molecular ratio extracting from the fitting results of SAXS profiles in (a) by Scatter software.

From fitting the SAXS profiles, it is revealed that the PS-b-PNIPAM DBC is forming lamellar structure in which the $q/q*$ ratio is 1, 2 and 3. The inter-lamellar distance of the lamellar structure can be obtained from the position of the first and most intense scattering peak using the formula $q=2\pi/d$ as well as from the fitting of the SAXS profile. At small initial incorporation of iron salt for [Fe]/[NIPAM] ratio sample, it can be easily overserved that there is a strong shift to a small q value. The shift of main first scattering peak corresponds to an increasing of lamellar distance from 26 nm to 31.4 nm upon salt incorporation.

It is assumed that the iron ions reside in the PNIPAM chain due to its polarity, forming a complex with the polar groups on the PNIPAM chain. The phenomenon for this lamellar expansion can be interpreted by the order-disorder transition (ODT) theory, which is consist of three parameters, $\chi$-the incompatibility of A and B block, N-degree of polymerization and $f_A$-volume fraction of A component. According to the Semenov Method, the lamellar distance $d$ is determined by a balance between interfacial energy and the energy of stretching the blocks of the copolymer, and the given formula is (5-1): [39]

$$d \sim a N^{\frac{2}{3}} \chi^{\frac{1}{6}} \tag{5-1}$$

Where $a$ is the size of a monomer. Thus the increasing of lamellar distance is due to the increasing of the incompatibility $\chi$ of PS and PNIPAM block. This is also confirmed by observing pronounce tertiary peak in the SAXS profile at [Fe]/[NIPAM] = 0.01 while in the SAXS profile of pure PS-b-NIPAM the lamellar structure is less ordered (showing only two scattering peaks).

In addition, with the increasing of [Fe]/[NIPAM] molecular ratio, the pronounced lamellar structure starts to disappear upon further increase of [Fe]/[NIPAM] ratio > 0.01. At [Fe]/[NIPAM] ratio ≥0.1, the lamellar structure is completely vanished and the structure is mainly ill-defined.

## 5.1.2 Domain Orientation

The lamellar domain orientation of the <u>PS</u>-b-PNIPAM hybrid nanocomposites seems to depend on the iron oxide content inside the polymer. In figure 5-2 (a) the SAXS 2D patterns of iron oxide/ <u>PS</u>-b-PNIPAM hybrid nanocomposites for different [Fe]/[NIPAM] ratio are presented. And in figure 5-2 (b) the intensity of the secondary scattering ring are plotted using Scatter software.

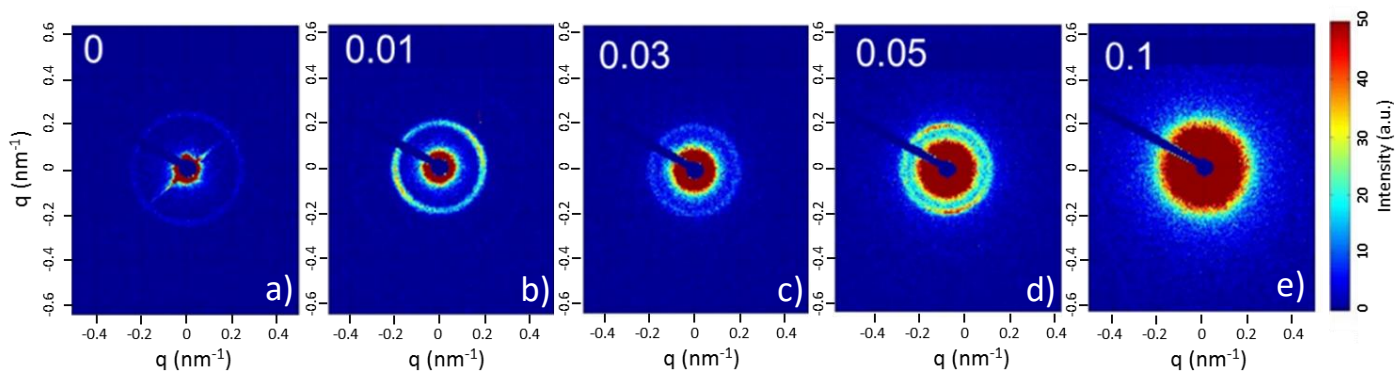

Figure 5-2: SAXS 2D profiles of iron salt / <u>PS</u>-b-PNIPAM DBC nanocomposites with [Fe]/[NIPAM] molecular ratio of 0, 0.01, 0.03, 0.05 and 0.1 corresponding to figure a) to e).

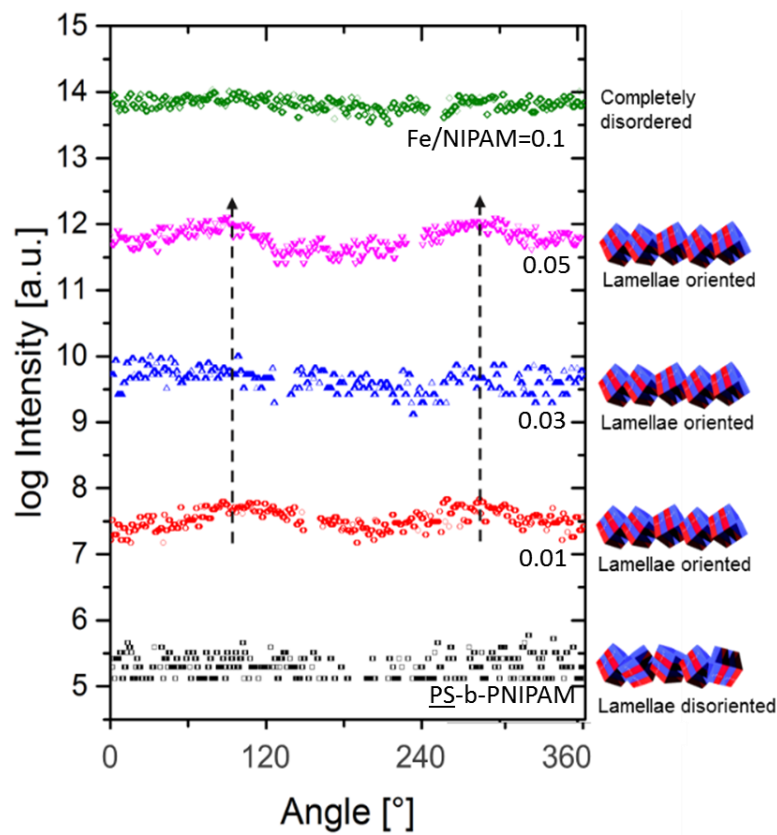

Figure 5-3: Intensity plot of the secondary scattering ring of the SAXS 2D profiles for iron oxide / <u>PS</u>-b-PNIPAM DBC nanocomposites with [Fe] /[NIPAM] molecular ratio of 0, 0.01, 0.03, 0.05 and 0.1. The right sketches show the corresponding orientation states of the lamellar domains.

Apparently the orientation of the lamellar domain can be observed in figure 5-2 and figure 5-3, there is no preferential orientation for the bare DBC while the lamellar domain starts to be preferentially oriented upon adding iron salt. At high [Fe]/[NIPAM] ratio both the lamellar structure and the orientation are vanishing thus there is no ordered structure anymore. The initial increase of preferential orientation indicates the possible accommodation of iron oxide within a specific domain, in this case the PNIPAM block. At high iron salt concentration, the formation of ill-defined structure indicates the loss of high selectivity of iron salt to one block. At [Fe]/[NIPAM] ⩾ 0.1, the PNIPAM chain is no longer able to fulfill the coordination requirement of the Fe ions.

## 5.1.3 Thermal Stability

The in situ SAXS measurements were performed on both thermally treated bare DBC and metal oxide/hybrid materials. The samples were stabilized for 1 hour at each temperature for SAXS measurement. The SAXS measurement results are shown in figure 5-4, and there is no peak position shift observed, which means the structures of both bare PS-b-PNIPAM DBC and bare PS-b-PNIPAM DBC at dry state are thermal stable until 175ºC.

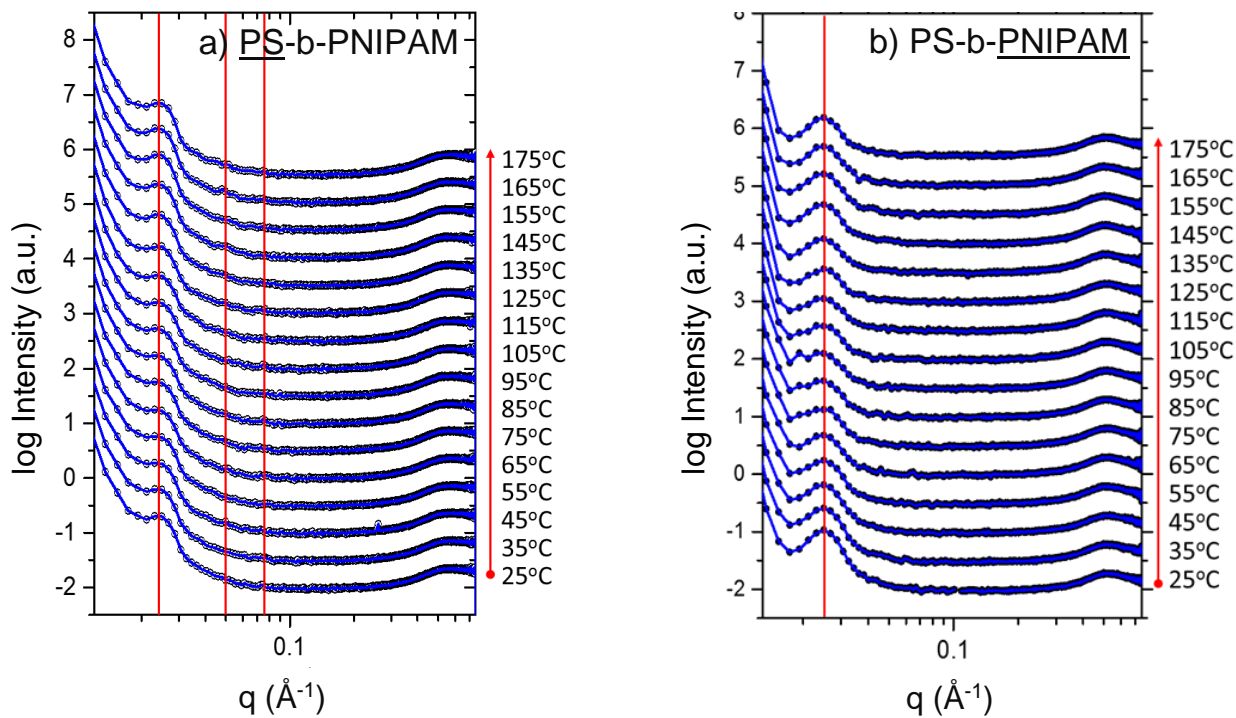

Figure 5-4: The SAXS profiles of the a) PS-b-PNIPAM and b) PS-b-PNIPAM bulk sample annealed at 130ºC for 48 hours, then heated from 25 ºC to 175 ºC in the steps of 10 ºC. The SAXS profiles are shifted along the y-axis for clarity purposes.

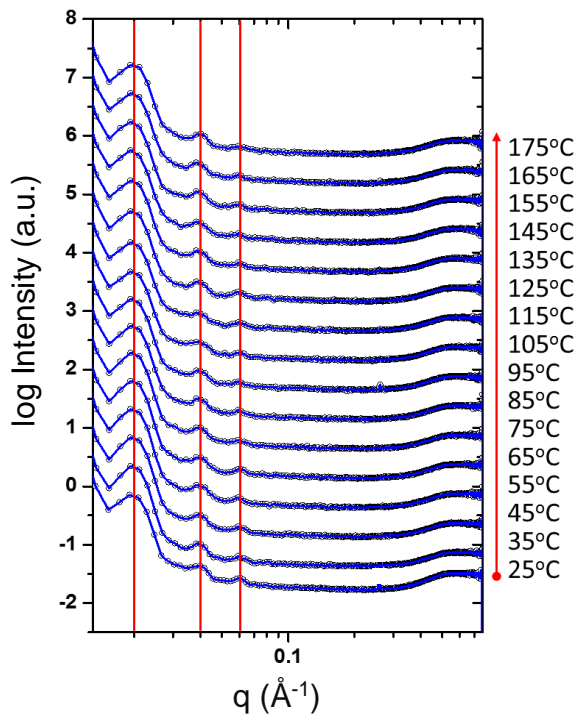

Figure 5-5: The SAXS profiles of the PS-b-PNIPAM / iron oxide hybrid bulk material with molecular ratio of [Fe]/[NIPAM]=0.01 annealed at 130ºC for 48 hours, then heated from 25 ºC to 175 ºC in the steps of 10 ºC. The SAXS profiles are shifted along the y-axis for clarity purposes.

The thermal stability of the PS-b-PNIPAM DBC /iron oxide hybrid nanocomposites with [Fe]/[NIPAM]=0.01 that prior-annealed at 130ºC for 48 hours, is also studied by SAXS measurements at different temperatures. The results presented in figure 5-5 show that the lamellar structure of PS-b-PNIPAM DBC/iron oxide hybrid materials does not modify even at high temperatures, indicating thermally stable nanocomposites up to 175ºC.

## 5.1.4 Surface Morphology

The surface morphology of the diblock copolymer films can be probed by scanning electron microscopy (SEM). The fitting results of the SAXS profile show that the structure of PS-b-PNIPAM is lamellar and for PS-b-PNIPAM is mixing of lamellar and cylinder structure. The SEM results show the block structure of PS-b-PNIPAM in a thin film format confirms the SAXS results. The film samples for both PS-b-PNIPAM and PS-b-PNIPAM bare DBCs are prepared according to the spin-coating method described in chapter 3 with 45 mg/ml DBC concentration and acid cleaned silicon substrates.

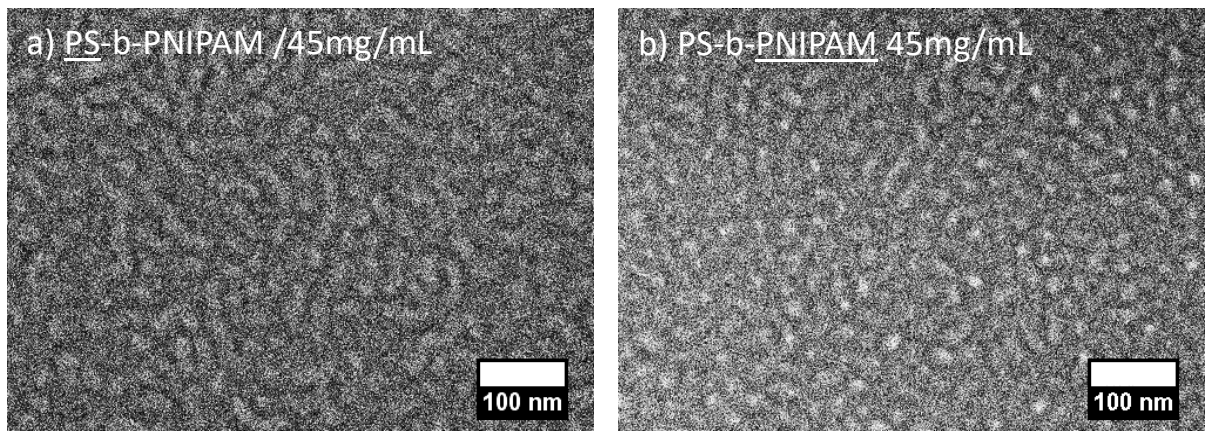

Figure 5-6: SEM images of both bare a) PS-b-PNIPAM and b) PS-b-PNIPAM DBCs.

Figure 5-6 shows that the structure of PS-b-PNIPAM DBC film is mostly parallel lamellar structure and for PS-b-PNIPAM DBC is lamellar mixed with vertical cylinder structure. These films were prepared by spin coating a DBC solution at concentration of 45 mg/mL. For low solution concentration of about 10 mg/mL, the as-prepared PS-b-PNIPAM DBC thin film shows mainly vertical cylinder structure (See figure A-1 in appendix).

# 5.2 Thermoresponsive Behavior of PS-b-PNIPAM DBC Bulky Films: In-situ and Ex-situ SAXS Study

In the liquid systems, the thermoresponsive behavior of PNIPAM homopolymer and BCs have been extensively investigated [2-5]. While the thermoresponsive behavior of nanostructured PNIPAM based BCs as a bulky material is barely reported. The possible water vapor swelling of PNIPAM domains in nanostructured PS-b-PNIPAM DBC systems opens the pathway to easily manipulate the morphology of these BCs.

## 5.2.1 In-situ SAXS study of PS-b-<u>PNIPAM</u> DBC bulky film

Here, in-situ SAXS investigation of free-standing PS-b-PNIPAM DBC bulky samples upon water vapor exposure is presented. The PNIPAM dominated PS-b-<u>PNIPAM</u> DBC means large volume fraction of PNIPAM block and referred to using underlined PNIPAM block. The free standing PS-b-<u>PNIPAM</u> bulky sample is prepared according to the method described in chapter 3, and it is mounted in the temperature and humidity controlling cell as described in chapter 4.

This in-situ SAXS experiment aims to observe the thermoresponsive behavior/morphology of PS-b-<u>PNIPAM</u> DBC, during swelling at low temperature below LCST and deswelling at high temperature above LCST, by controlling both the temperature and relative humidity (r.H). Since the relative humidity is not an independent parameter from temperature, it is calculated by the ratio of the partial pressure of water vapor ($p_{H2O}$) in the mixture to the equilibrium vapor pressure of water over a flat surface of pure water ($p_{*H2O}$) at a given temperature. Thus when the temperature is increased, the relative humidity decreases due to the calculation method while the absolute humidity (mg/L) keeps the same. The relative humidity has an important effect on the swelling and deswelling of PS-b-PNIPAM, thus the way to control the relative humidity at different temperature is further controlled by increasing or decreasing water content in the cell using water injection or bubbling. Each SAXS measurement is a scan of half an hour.

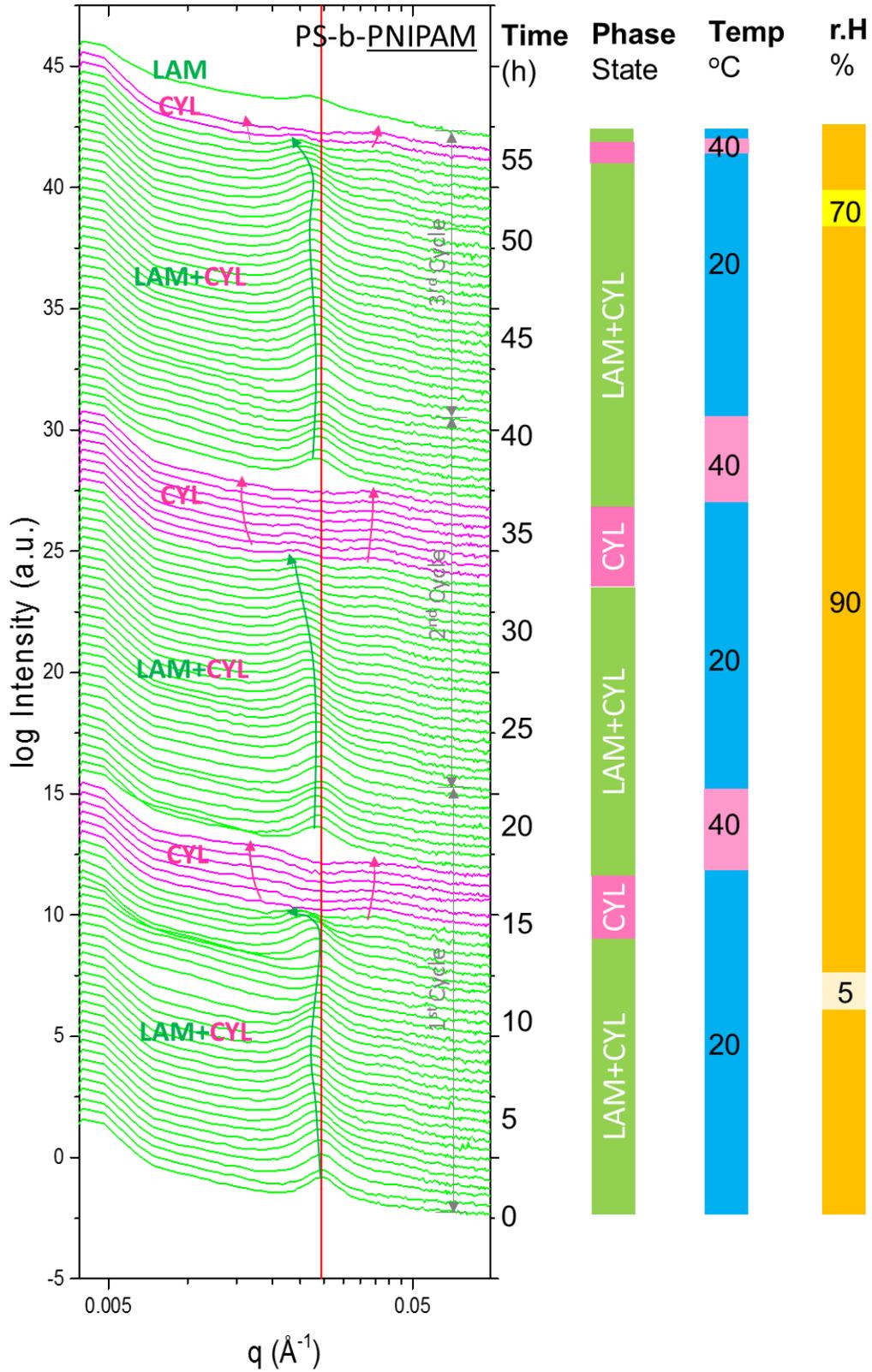

Figure 5-7: Evolution of in-situ 1D SAXS profiles of bare PS-b-PNIPAM DBC during swelling/de-swelling process. The red vertical line indicates the initial peak positon.

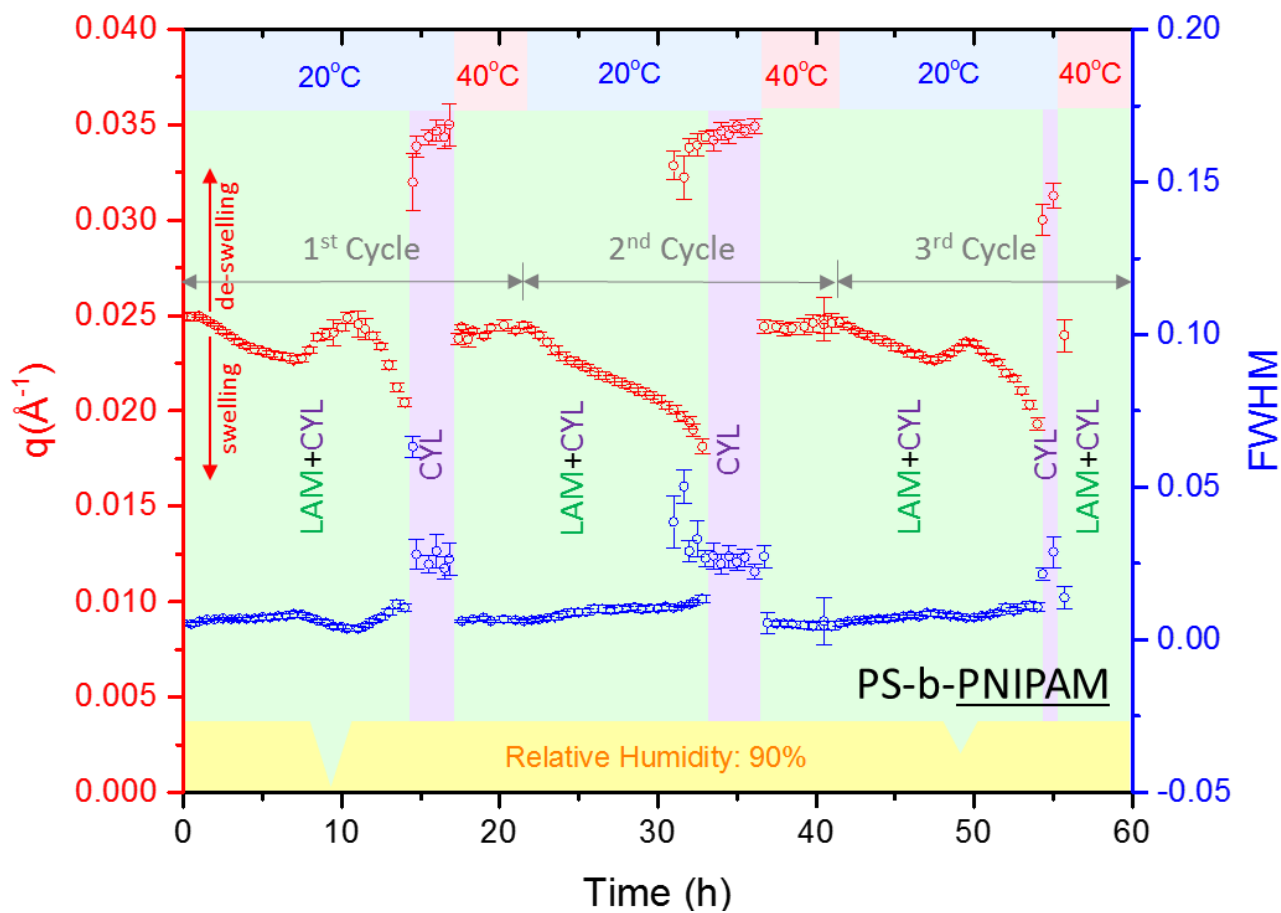

Figure 5-8: Peak fitting results of 1D SAXS profiles for $q$ position and FWHM and its evolution versus time during three temperature cycles.

To analyze the evolution of the phase transition in figure 5-7, the peak positions are fitted and the results are given in figure 5-8 with temperature and humidity conditions indicated. The impacts of relative humidity and temperature are discussed below respectively.

### a) Influence of relative humidity

The relative humidity dominates the water absorption process of the PS-b-PNIPAM system. Due to the amide group in NIPAM molecule functioning as hydrophilic group, the water can be either bonded chemically with NIPAM molecule by hydrogen bond or physically absorbed. As the increasing of water amount inside the polymerized NIPAM molecules, the molecules will expand to certain level then the expansion will stop and the system will reach an equilibrium state because of the balance of elastic shrinking force and water absorbing expansion.

There are three temperature cycles for this in-situ SAXS experiment and for each cycle, as seen in figure 5-7 and figure 5-8, during the first temperature cycle the q value decreases gradually at the beginning with temperature keeping constantly at 20ºC, which means the smoothly swelling of the PS-b-PNIPAM free standing film.

After about 7 h the relative humidity is decreased at a very low level (r. H=5%), it shows the deswelling of the system (figure 5-8), as indicated by reverse shift of the main characteristic q peak to higher values due to the decreasing of relative humidity. Interestingly, we can bring the system to almost its initial state as indicated by reaching the same initial q value. Three hours later, the relative humidity is back ramped up to 90% then the system is spontaneously re-swelled again indicating a fast response to the humidity environment at 20ºC.

In the second temperature cycle (Figure 5-8), the high relative humidity (r.H=90%) is kept for longer time and the results show a systematic decrease of the q values upon swelling of the system till a phase transition process occurs. Finally, in the third cycle, the humidity is operated at an intermediate level (r.H=70%) between 48 and 50 hours, and the swelling curve gives the same responsive behavior corresponding to the humidity change, but at a slower rate as indicated from the slopes of the time-dependent q values. Our experiment proves high sensitivity of the responsive block to the environment humidity level. This behavior has rarely investigated for the thermal responsive polymer films, where temperature is only switched at certain humidity level to investigate the swelling/deswelling behavior.

The most important thing is that the high relative humidity seems to initiate a phase transition from lamellar mixed cylinder structure to cylinder structure. This can be observed in all of the three cycles in figure 5-8 and it means the thermoresponsive behavior of the PS-b-PNIPAM shows reversible behavior and the free standing film itself is quite stable. The phase transition occurs when the main peak referring to lamellar mixed cylinder structure disappears and two new peaks arise referring to new cylinder structure. Due to the expansion of PNIPAM block in the high relative humidity while PS block keeping rigid, the volume ratio of PNIPAM block increases, which gives new block structure as cylinder.

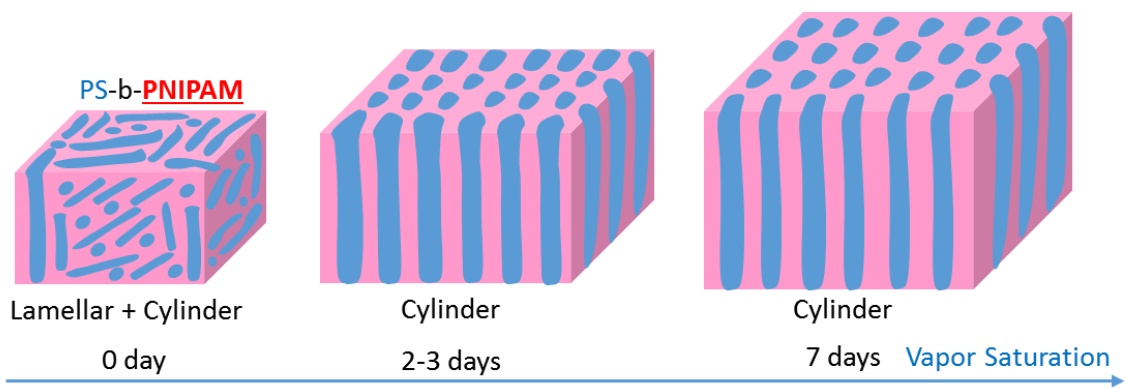

Figure 5-9: Schematic of block structure evolution versus time for PS-b-PNIPAM in high relative humidity.

## b) Influence of temperature

The temperature plays a very import role during the thermoresponsive behavior of PS-b-PNIPAM. As is shown in figure 5-8, in the first swelling cycle, a phase transition occurs, the system is kept for several hours, then the temperature is increased from 20ºC to 40ºC. Due to temperature jump, the PNIPAM block dramatically shrinks and deswell most of the physically absorbed water. This is indicated by the dramatic shift of the q values (within 30 min measurement time) to its initial value. A second phase transition induced by temperature jump is indicated in concomitant with the fast de-swelling behavior of the system.

Equilibrating the system at 40ºC for few hours, the q value is slightly increased until the film is under equilibrium. This oscillation behavior of the system is related to the relaxation of the chain upon water removal. This behavior is previously observed by Q. Zhong et al [40] on homopolymer hydrogel systems. In the second and third cycle, the free standing film shows similar thermoresponsive behavior indicating mechanically stable film. The stabilities of the mechanical structure and thermoresponsive property are significantly important for some applications such as sensors and actuators.

## 5.2.2 In-situ SAXS study of PS-b-PNIPAM DBC bulky film

The PS-b-PNIPAM is PS dominated diblock copolymer DBC which means the PNIPAM is in the PS matrix. Using the static SAXS and SEM investigation, the lamellar structure of the PS-b-PNIPAM DBC is proved. In principle the PNIPAM block should be able to expand under a high relative humidity circumstances, but since both chain ends are confined by the PS block (glassy rigid block), the swelling behaviour of this DBC should be significantly limited. The swelling/deswelling behaviour of this DBC with major PS block compared with PS-b-PNIPAM DBC with a major PNIPAM block is an interesting topic to study.

The PS-b-PNIPAM bulk sample is prepared according to the method described in chapter 3 with mica window as the solution casting substrate, and it is positioned in the temperature and humidity controlling cell as described in chapter 4. The in-situ SAXS study was performed for 30 hours with each SAXS measurement scans of half an hour.

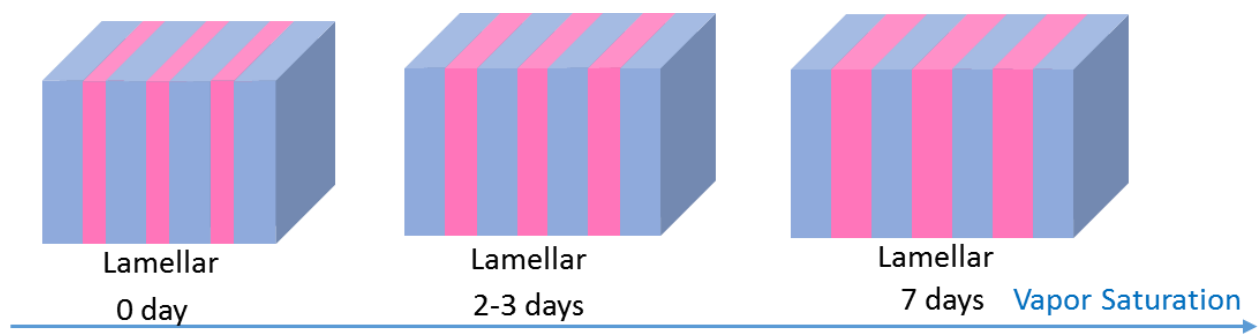

Figure 5-10: Schematic of block structure evolution versus time for PS-b-PNIPAM in high relative humidity.

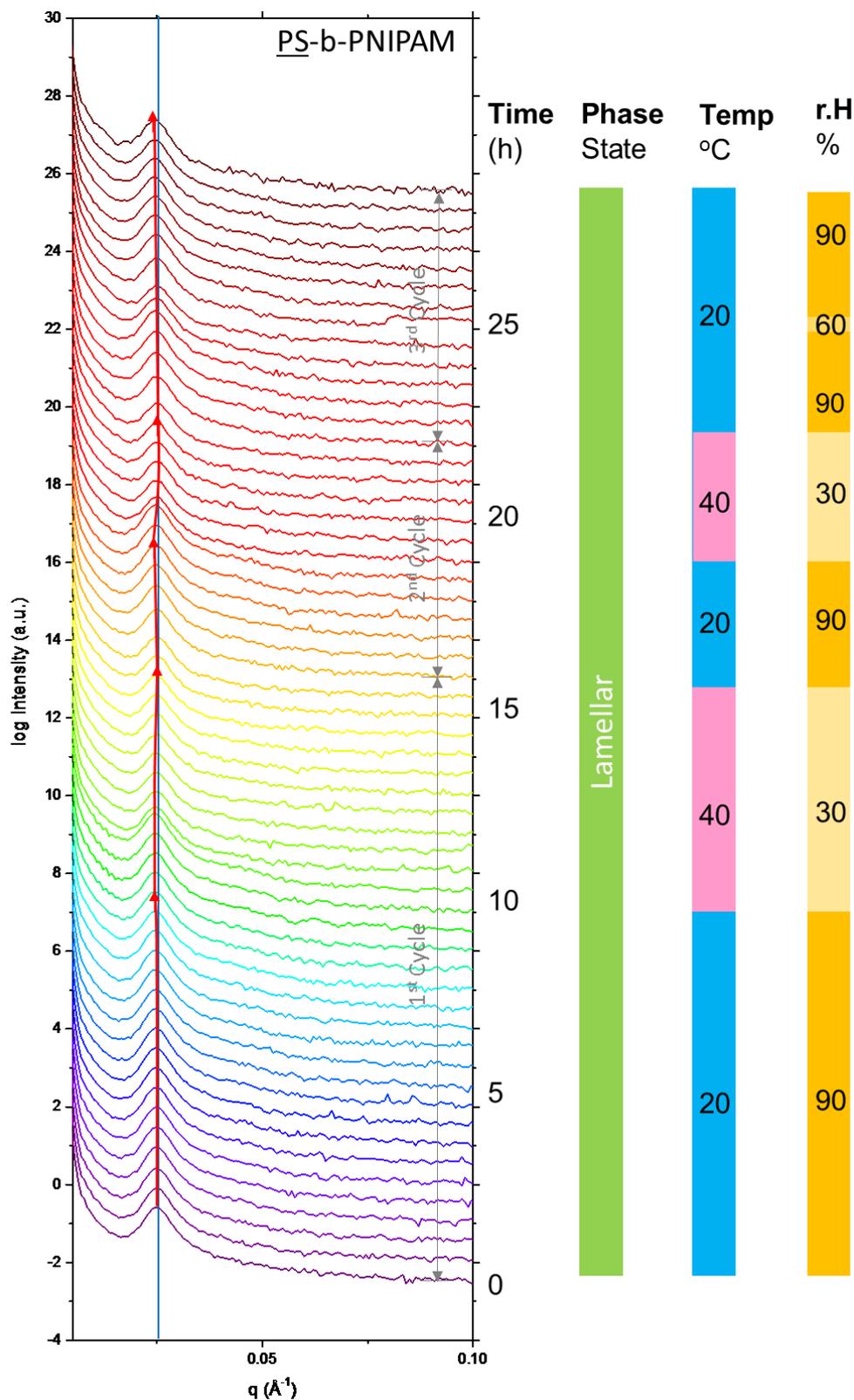

Figure 5-11: Evolution of in-situ 1D SAXS profiles for bare PS-b-PNIPAM DBC during swelling/deswelling process. The blue vertical line indicates the initial peak positon.

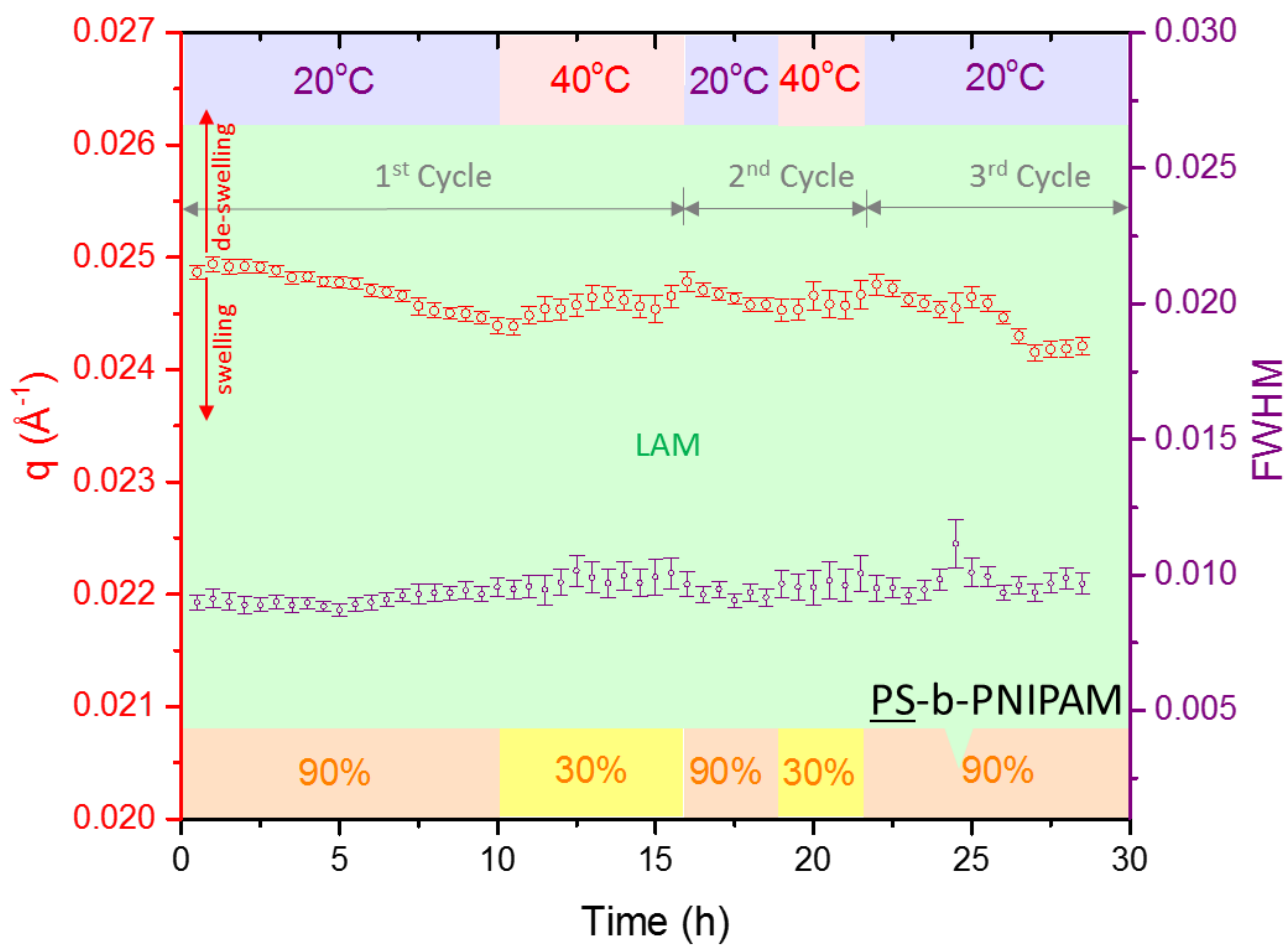

Figure 5-12: Peak fitting results of 1D SAXS profiles for q position and FWHM and its evolution versus time during three temperature cycles.

The in-situ SAXS results for PS-b-PNIPAM bulky film show that it undergoes very limited volume expansion but much slower than PS-b-PNIPAM bulky film. During the first temperature cycle, due to the controlling of the humidity is based on the absolute humidity, the relative humidity drops when the temperature increases. Both of the increasing temperature and decreasing of relative humidity contribute to the deswelling of PNIPAM block, thus the q value goes back to near the initial value. And there is no phase transition during the expansion, which is because of the confinement of the rigid glassy PS block. During the second cycle and third half cycle, the behavior of the film is similar to that of the first cycle. This also confirms that the PS-b-PNIPAM DBC bulky film is stable through thermoresponsive behavior. In principle, our study proves the assumption that the PS-b-PNIPAM DBC with major glassy PS block forms a rigid matrix limiting the flexibility of the PNIPAM block to freely expand upon swelling, indicating a very limited responsive behavior compared with PS-b-<u>PNIPAM</u> DBC with PNIPAM major block. Similarly, this has been recently observed by Q. Zhong et al [40] where it is observed using neutron reflectivity technique that PS layer within nanoscale multilayered PS/PNIPAM stack structures hinders further water absorption in buried layers.

### 5.2.3 Ex-situ SAXS study of PS-b-PNIPAM DBC bulky film

During the in-situ SAXS experiments, both the <u>PS</u>-b-PNIPAM free standing film and PS-b-<u>PNIPAM</u> bulk sample didn't reach a final equilibrium states at high relative humidity. Thus the ex-situ SAXS experiments were done for exploring how far the peak can shift in a quasi-final equilibrium state (1-2 week swelling experiment).

A small plastic sample box with cap is used and a small reservoir made by bottle cap is built inside for generating of vapor atmosphere. Then the prepared DBC bulk samples are put in and the water or water/THF mixed solution are injected in the reservoir. The sample boxes are closed and sealed with taps and stored in a stable place for 1 ~ 2 weeks.

- **PS-b-PNIPAM DBC bulky film: ex-situ SAXS study**

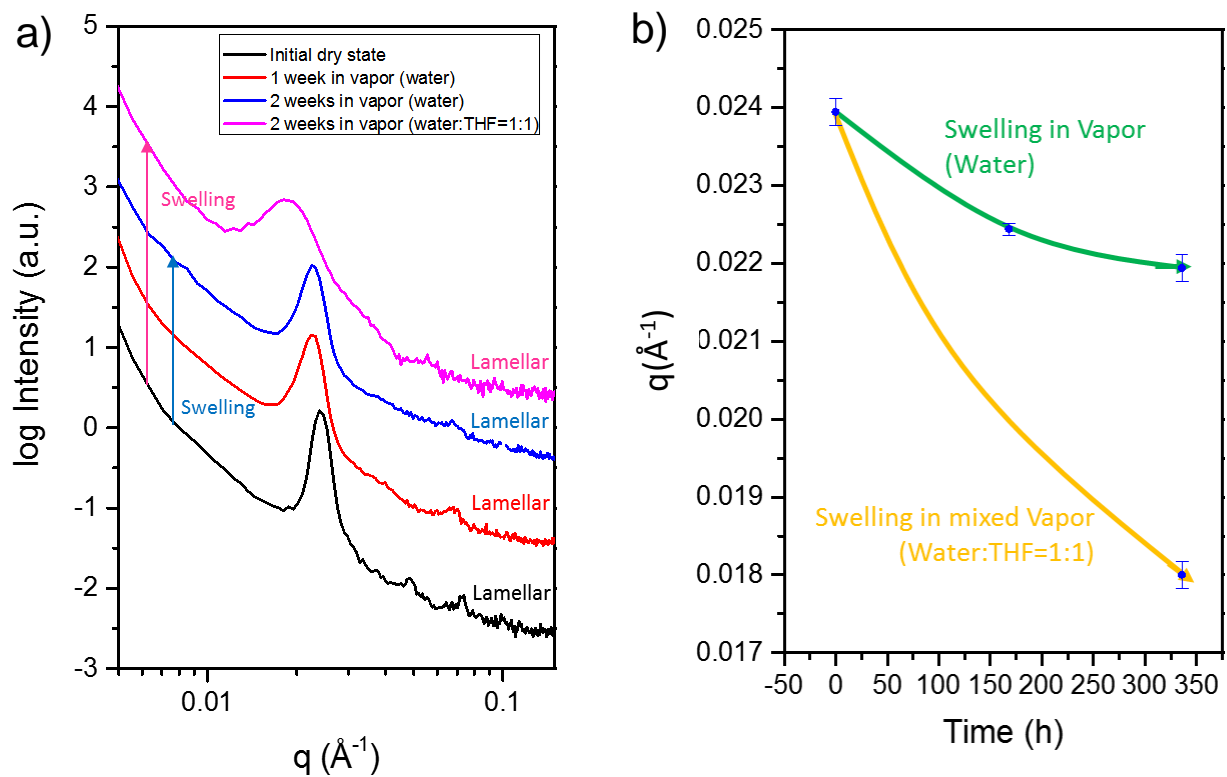

Figure 5-13: a) SAXS 1D profiles of PS-b-PNIPAM DBC bulk sample in water vapor and water:THF=1:1 mixed vapor for 1~2 weeks. b) Peak fitting results of the 1D SAXS profiles in figure a).

In figure 5-13, the ex-situ SAXS experiment results show that for PS-b-PNIPAM bulk sample in the water vapor, the shift of main peak q position stops after 2 weeks. But for the PS-b-PNIPAM bulk sample in the water:THF=1:1 mixed vapor, the main peak goes far from the original position and the peak also becomes more broad. This happens because both PS and PNIPAM have good solubility in THF, thus the THF vapor enables PS matrix to be more flexible which allows PNIPAM to expand more. But at the same time THF has the mixing effect on PS and PNIPAM block, thus the microphase separation process is affected and weak segregation process is observed as indicated from the peak broadening.

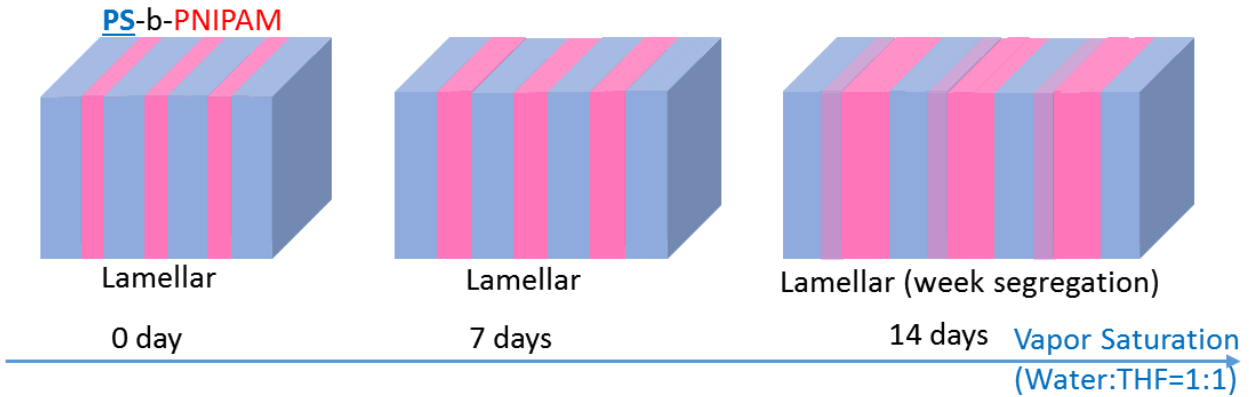

Figure 5-14: Schematic of block structure evolution versus time for PS-b-PNIPAM in water/THF=1:1 mixed vapor.

- **PS-b-PNIPAM DBC bulky film: ex-situ SAXS study**

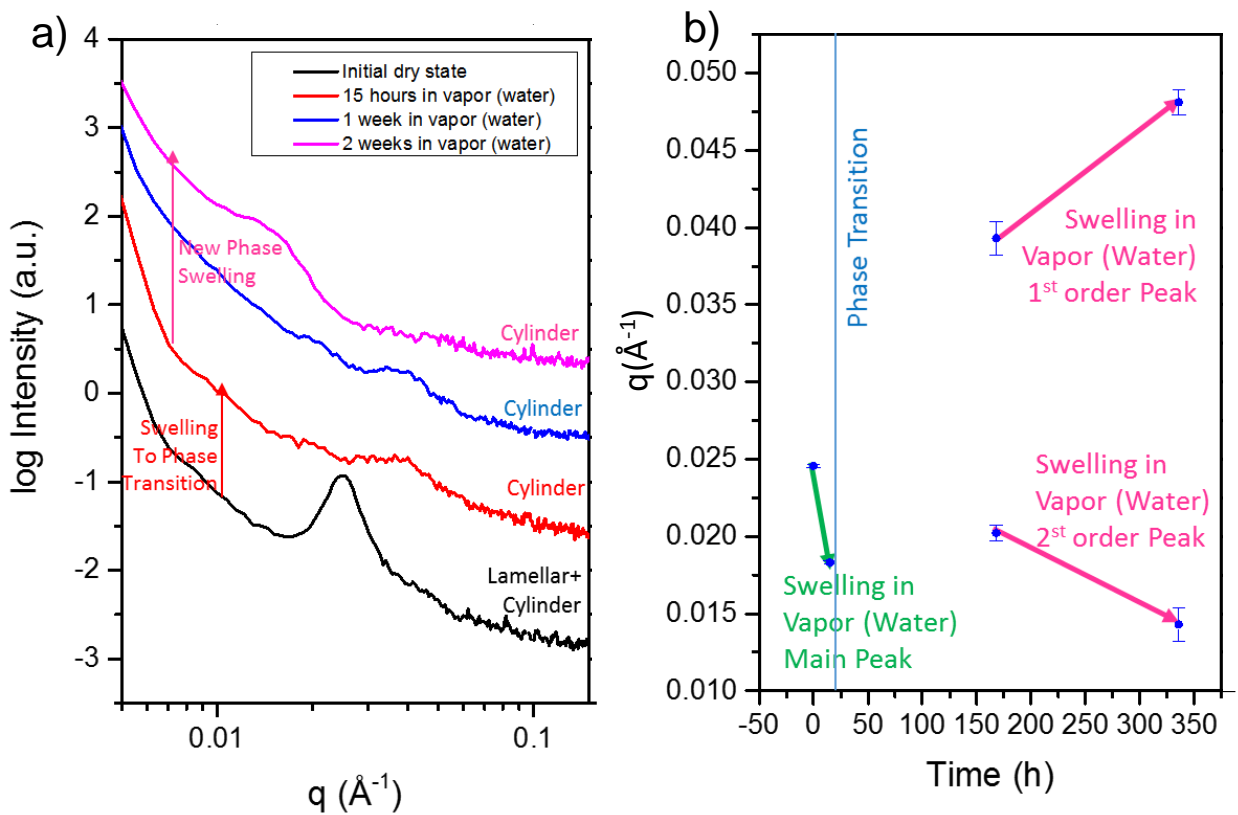

Figure 5-15: a) SAXS 1D profiles of PS-b-PNIPAM DBC bulk sample in water vapor for 2 weeks. b) Peak fitting results of the SAXS profiles in figure a).

The ex-situ SAXS experiments for PS-b-PNIPAM DBC bulk sample were performed with the same set-up as PS-b-PNIPAM, the sample is prepared as a free standing film. As shown in figure 5-15 (a), the swelling of PNIPAM brings the block structure to a phase transition process (as previously observed in the in-situ SAXS experiment, figure 5-8) and after two weeks the new phase peaks are still slightly shifting and growing. The equilibrium state seems need even long time to achieve as it can be estimated in according to figure 5-15 (b). This is generally an interesting behavior that the film adapts more and more water content but this can be also coupled with a possible deterioration of the free-standing film upon long water vapor exposure. It has been previously observed by Q. Zhong et al [41] that thermoresponsive hydrogel thin film upon swelling may de-wet upon longer time exposure to high-humidity environment. The robustness of the free-standing film upon very long time high humidity exposure to prepare integer and stable devices can be further examined.

## 5.3 Magnetic Behavior of PS-b-PNIPAM DBC / Iron Oxide Hybrid Bulky Films: SQUID Measurements

The magnetic properties of the PS-b-PNIPAM DBC /iron oxide hybrid nanocomposites bulky films (solution casting) are investigated by using a sensitive superconducting quantum interference device (SQUID) magnetometer. The magnetization measurements are performed with external magnetic field from -700 mT to 700 mT applied along the film plane. The B-M plot are presented in figure 5-16.

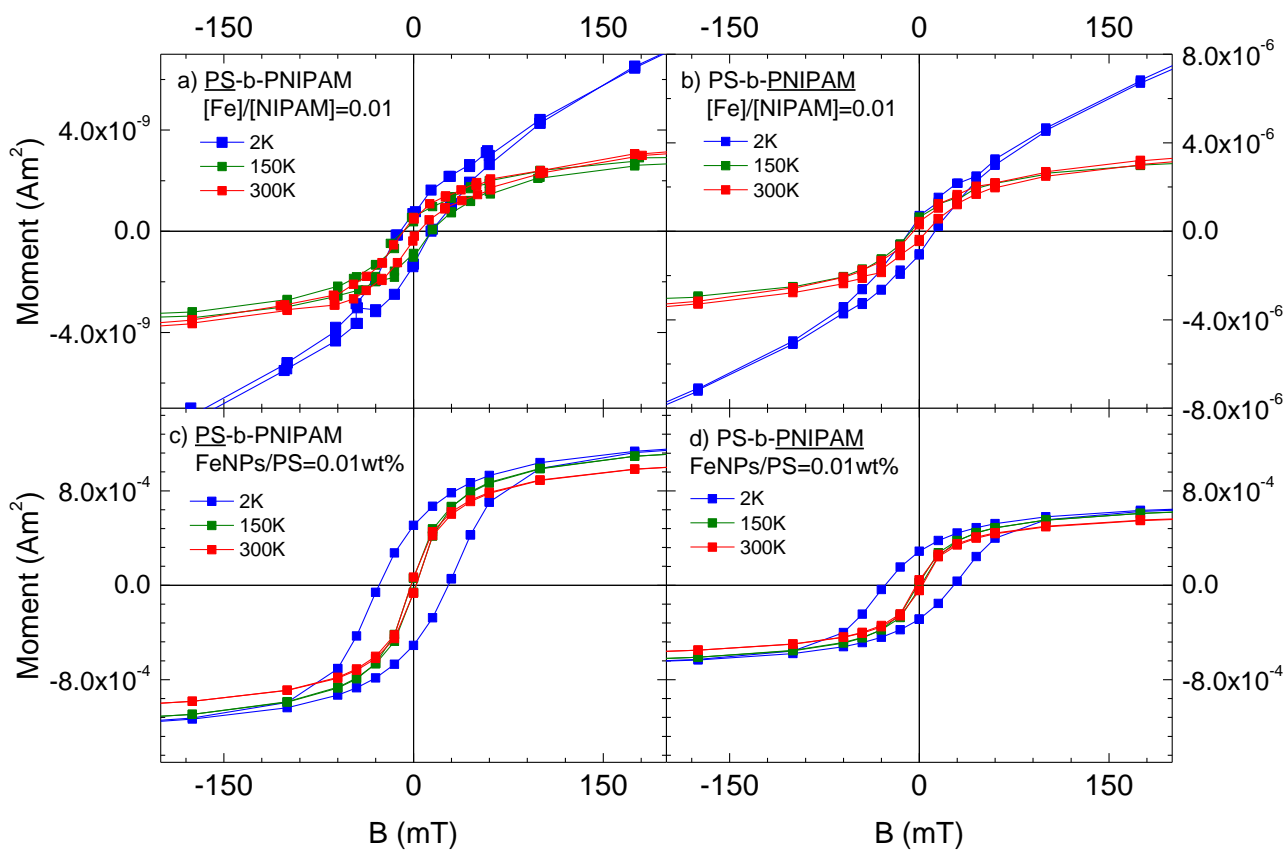

Figure 5-16: Magnetic moments measured as a function of external magnetic fields at 2K, 150K and 300K for PS-b-PNIPAM DBC / Fe salt or Fe NPs nanocomposites: a) PS-b-PNIPAM/Fe Salt ([Fe]/[NIPAM]=0.01); b) PS-b-PNIPAM/Fe Salt ([Fe]/[NIPAM]=0.01); c) PS-b-PNIPAM/Fe₂O₃ NPs (FeNPs/PS=0.01wt%); d) PS-b-PNIPAM/Fe₂O₃ NPs (FeNPs/PS=0.01wt%).

The magnetization measurements show that in figure 5-16 (a) and (b), the PS-b-PNIPAM DBC / Fe salt nanocomposites have strong hysteresis loops at low temperature while at high temperature the hysteresis loops become narrow and near to zero, which is the fingerprint for superparamagnetic materials. For the PS-b-PNIPAM DBC / $Fe_2O_3$ NPs nanocomposites, as the temperature increasing, they also behave as superparamagnetic material [13].

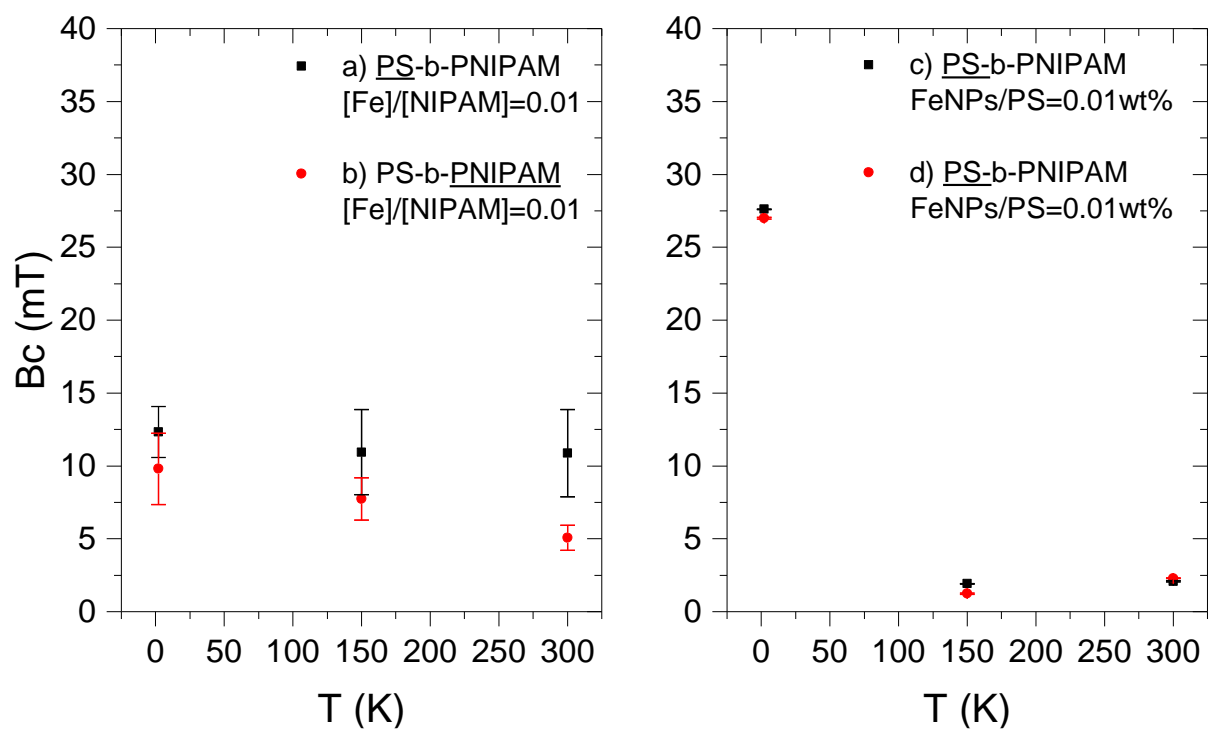

Figure 5-17: Coercivity $B_c$ extracted from the measured magnetic moments at different temperatures 2K, 150K and 300K for PS-b-PNIPAM DBC / Fe salt or Fe NPs nanocomposites: a) PS-b-PNIPAM/Fe Salt ([Fe]/[NIPAM]=0.01); b) PS-b-PNIPAM/Fe Salt ([Fe]/[NIPAM]=0.01); c) PS-b-PNIPAM/$Fe_2O_3$ NPs (FeNPs/PS=0.01wt%); d) PS-b-PNIPAM/$Fe_2O_3$ NPs (FeNPs/PS=0.01wt%).

The temperature-dependent magnetic behavior is shown for PS-b-PNIPAM DBC / Fe salt or Fe NPs nanocomposites. It can be observed that the hysteresis loop decreases with increasing temperature, which is proofed by previous studies [42-44]. For measurement at a relatively high temperature of 300K, the external magnetic fields are not enough to saturate magnetization. This is coincided with Curie's law [45]. The coercivity dramatically decreases with the temperature increasing and becomes near to zero at high temperature. This is the evidence for superparamagnetic behavior [13].

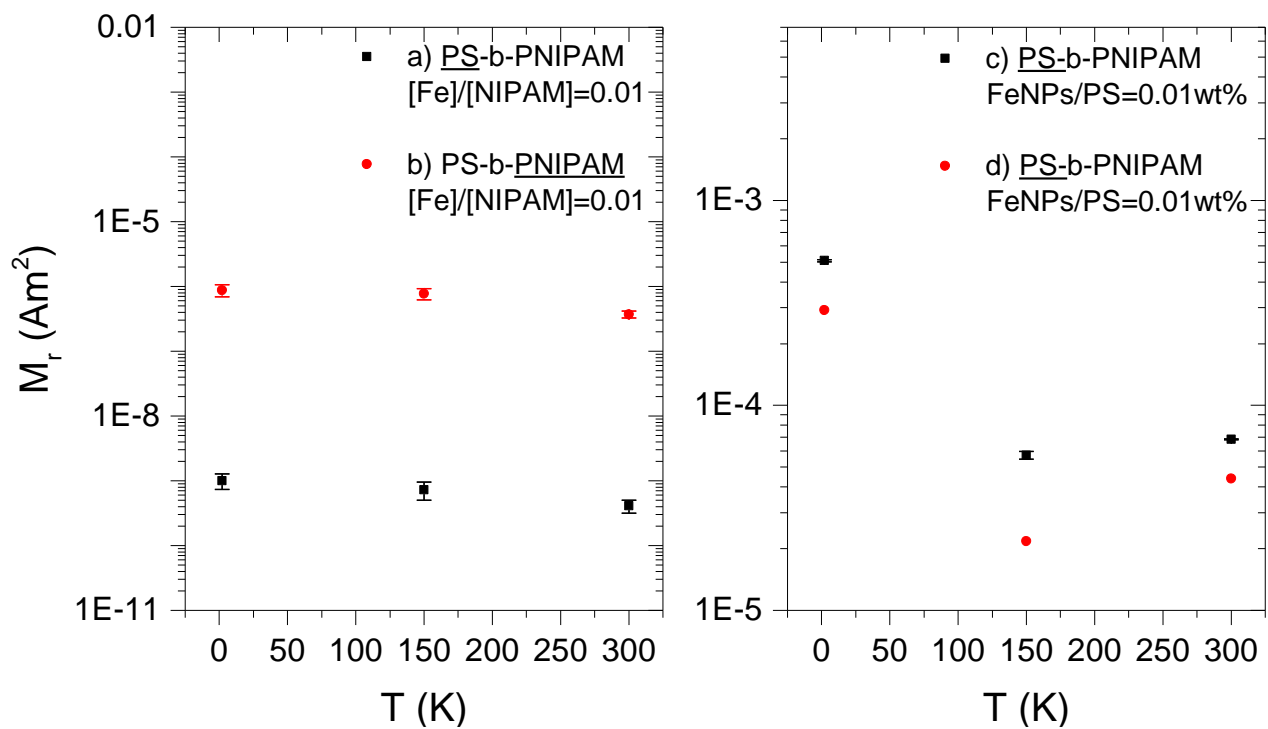

Figure 5-18: Remanence $M_r$ extracted from the measured magnetic moments at different temperatures 2K, 150K and 300K for PS-b-PNIPAM DBC / Fe salt or Fe NPs nanocomposites: a) PS-b-PNIPAM/Fe Salt ([Fe]/[NIPAM]=0.01); b) PS-b-PNIPAM/Fe Salt ([Fe]/[NIPAM]=0.01); c) PS-b-PNIPAM/Fe$_2$O$_3$ NPs (FeNPs/PS=0.01wt%); d) PS-b-PNIPAM/Fe$_2$O$_3$ NPs (FeNPs/PS=0.01wt%).

The remanence shows in figure 5-18 that it decays with the increasing of temperature as well. Both the temperature-dependent decay behaviors of coercivity and remanence are attributed to thermal fluctuations [13].

# 5.4 Structural Characterizations of PS-b-PNIPAM DBC / Iron Oxide Hybrid Thin Films: Static and Sputtering GISAXS Study

### 5.4.1 Static GISAXS study of PS-b-PNIPAM DBC/iron oxide hybrid thin film

The thin films of PS-b-PNIPAM DBC / iron oxide nanocomposite with increasing content of PS tailored $Fe_2O_3$ nanoparticles are prepared according to the method in chapter 3, and ex-situ GISAXS experiments were performed at SAXS beamline of Elettra Sincrotrone Trieste, Italy. The results are presented in figure 5-19 (a) and (b).

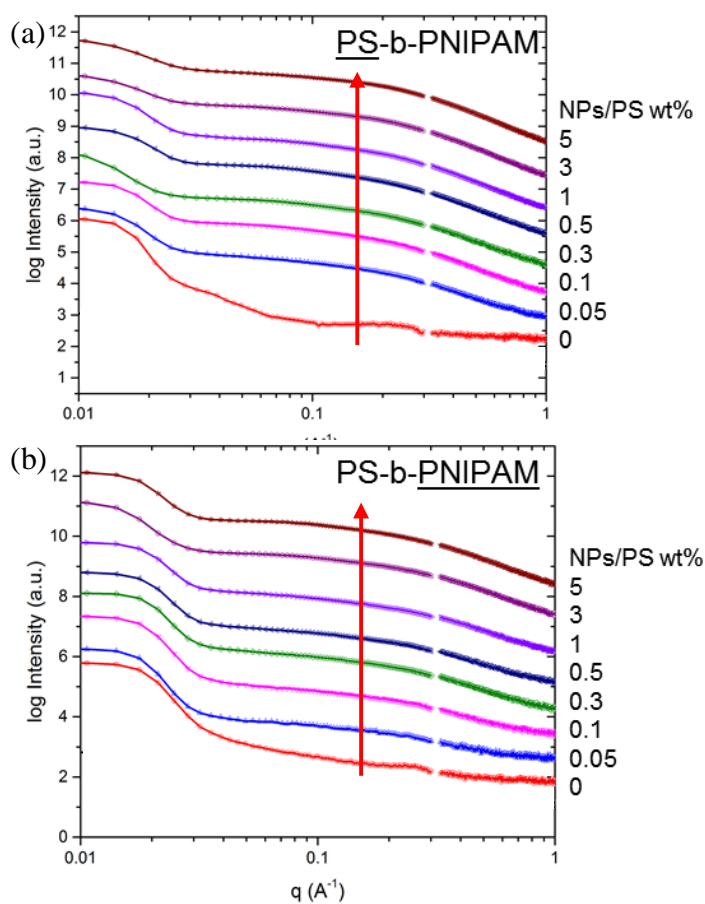

Figure 5-19: (a) The GISAXS $q_y$ profiles of the PS-b-PNIPAM / iron oxide nanocomposites with $Fe_2O_3$ nanoparticles weight ratio from 0 to 5 wt%. The GISAXS profiles are stacked on the intensity scale for comparison.

Figure 5-19: (b) The GISAXS $q_y$ profiles of the PS-b-PNIPAM / iron oxide nanocomposites with $Fe_2O_3$ nanoparticles weight ratio from 0 to 5 wt%. The GISAXS profiles are stacked on the intensity scale for comparison.

The bare diblock copolymer structure is relatively ill-defined and the $Fe_2O_3$ nanoparticles give a very wide peak which does not change with increasing its contents. The ill-defined structure of the DBC thin film compared with bulk samples is mainly due to surface physical/chemical effect on the evolved DBC structures. Possibly further optimization of the surface chemistry and film thickness is essential for further examinations. As we learned from the SAXS profile fitting results that the structure of bare PS-b-PNIPAM DBC bulk samples is lamellar with inter-lamellar distance equals to 26 nm. And the PS-coated $Fe_2O_3$ nanoparticles used here is around 10 nm, thus there is possible large confinement effect to incorporate these large size NPs. Possible thin film optimization of this PNIPAM-based DBCs and smaller sized metal-oxide NPs may allow formation of nanostructured hybrid responsive systems in thin film format. This part has not been further investigated in this thesis. However, similar in-situ scattering investigation on PNIPAM based nanostructured DBCs in thin film format to those performed for free-standing bulky films (in this thesis) is an interesting topic to further investigate.

## 5.4.2 Sputtering GISAXS study of bare PS-b-PNIPAM DBC thin film

The sputtering GISAXS experiments were performed at P03 beamline in DESY, Hamburg. The gold nanoparticles are sputtered on the surface of bare PS-b-PNIPAM DBC thin film with a rate of 0.007 nm/s and sputter power of 150 W, and the GISAXS measurements are operated at the same time to probe the structure change of the deposited gold layer and the growth kinetics of the gold nanoparticles.

From figure 5-20, it is clearly to see the growth process of the gold NPs. Around 3 min after the start of sputtering, the first gold metal characteristic peak appears as a cloud and a quasi-uniform gold layer around 15 min is formed, during this time, also the second and third ordered gold characteristic peaks come from high q values toward a low q values.

The Yoneda peak of the PS-b-PNIPAM is a material-specific characteristic, which reveals the strongest scattering position depending on the critical angle of the material.

$$(5\text{-}2)$$

$$Y \text{ (in pixel)} = y + \frac{d}{p}\tan[(\alpha_i + \alpha_c]$$

Where Y is the Yoneda peak position (in pixel) of the material, and y is the pixel position of the direct beam. d is the distance between sample and detector, and p is the size of the pixel. $\alpha_i$ is the incident angle of the x-ray, and $\alpha_c$ is the critical angle of the investigated material.

Thus the Yoneda peak position can be calculated for PS and PNIPAM, it is $q_z$= 0.567 nm$^{-1}$ (245.5 in pixel) and $q_z$= 0.576 nm$^{-1}$ (248.19 in pixel) respectively. And the 2D time-resolved mapping (figure 5-20) shows intensity evolution of the horizontal cuts at the Yoneda peak position of PS-b-PNIPAM.

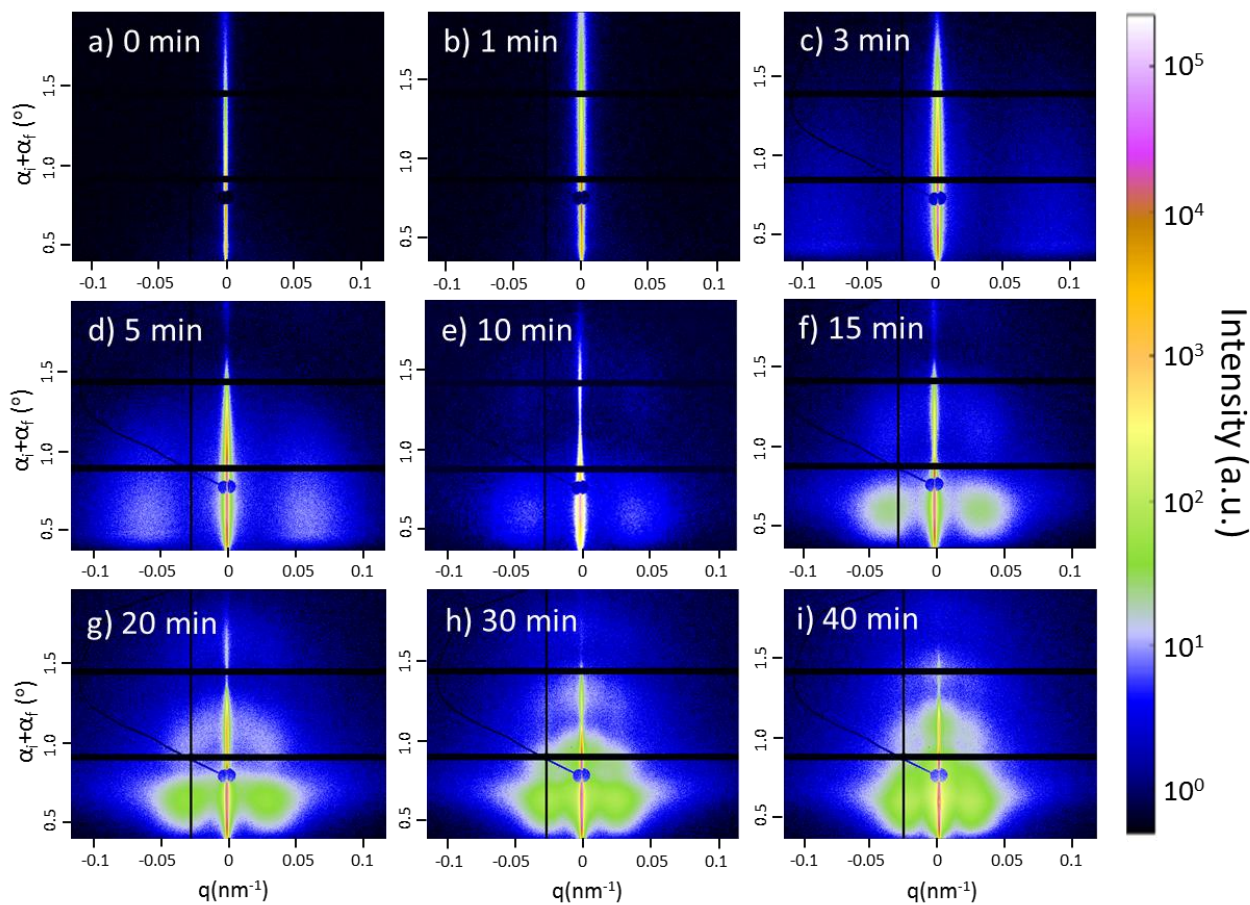

Figure 5-20: Time evolution of GISAXS 2D intensity patterns from a) 0 min to i) 40 min for bare PS-b-PNIPAM thin film during Au nanoparticles sputtering process.

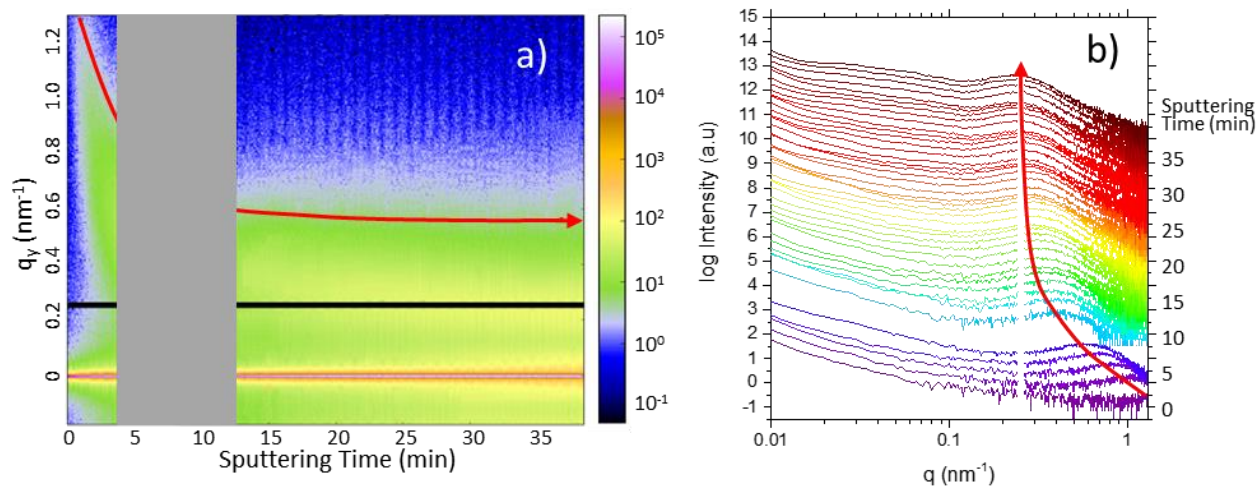

Figure 5-21: a) 2D time-mapping showing the temporal intensity evolution of <u>PS</u>-b-PNIPAM DBC bare thin film GISAXS horizontal cuts at $q_z = 0.567 \sim 0.576$ nm$^{-1}$. The grey area is covered due to an accidently beam dump, and it is the same reason for all the following graphs. b) Intensity evolution of PS-b-PNIPAM DBC GISAXS horizontal cuts at $q_z = 0.567 \sim 0.576$ nm$^{-1}$ with stacking on intensity scale for comparison.

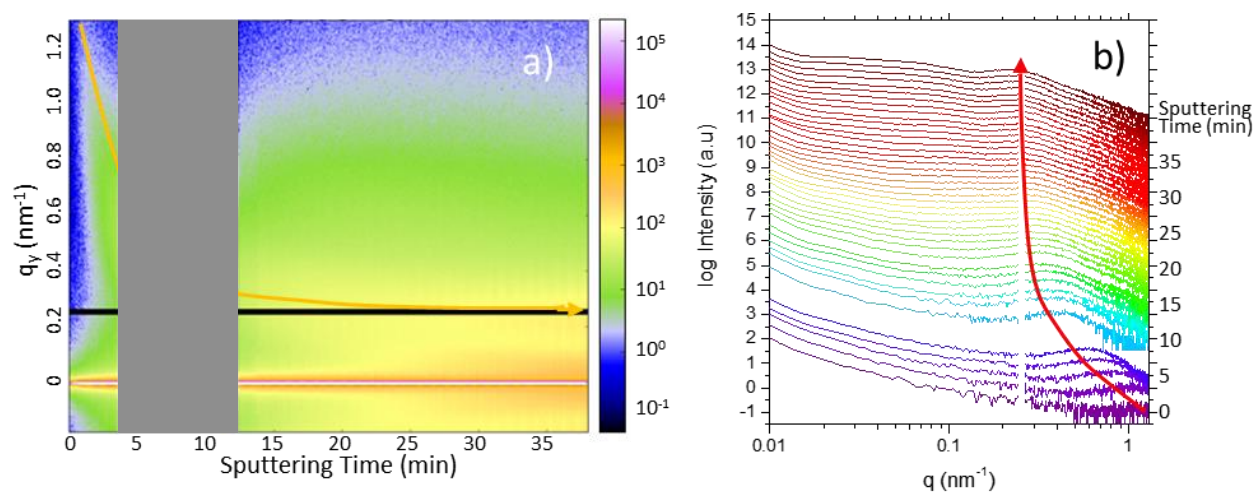

Figure 5-22: a) 2D time-resolved map showing the temporal intensity evolution of Au GISAXS horizontal cuts at $q_z = 0.851$ nm$^{-1}$. b) Intensity evolution of Au GISAXS horizontal cuts at $q_z = 0.861$ nm$^{-1}$ with stacking on intensity scale for comparison.

The same calculation can be applied for gold NPs, the Yoneda peak position of Au is given as $q_z = 0.851$ nm$^{-1}$ (345 in pixel). The time evolution of the gold scattering peak can be seen from figure 5-22 (b), it undergoes exponentially increasing in the early stage and finally stopped after sputtering for 30 min. The gold layer formation can be interpreted from nucleation, diffusion, adsorption to grain growth. In the very beginning, the gold nanoparticles exist as small particles and they are the nucleation center. The gold particles grow soon into gold cluster in the diffusion phase. Then these gold clusters adsorb and merge together until percolation threshold. Afterwards a dominated cluster grows by a movement of grain boundaries [46]. The gold layer thickness, however, will still increase and it forms multi-layer structure as shown in figure 5-23 (a) and (b).

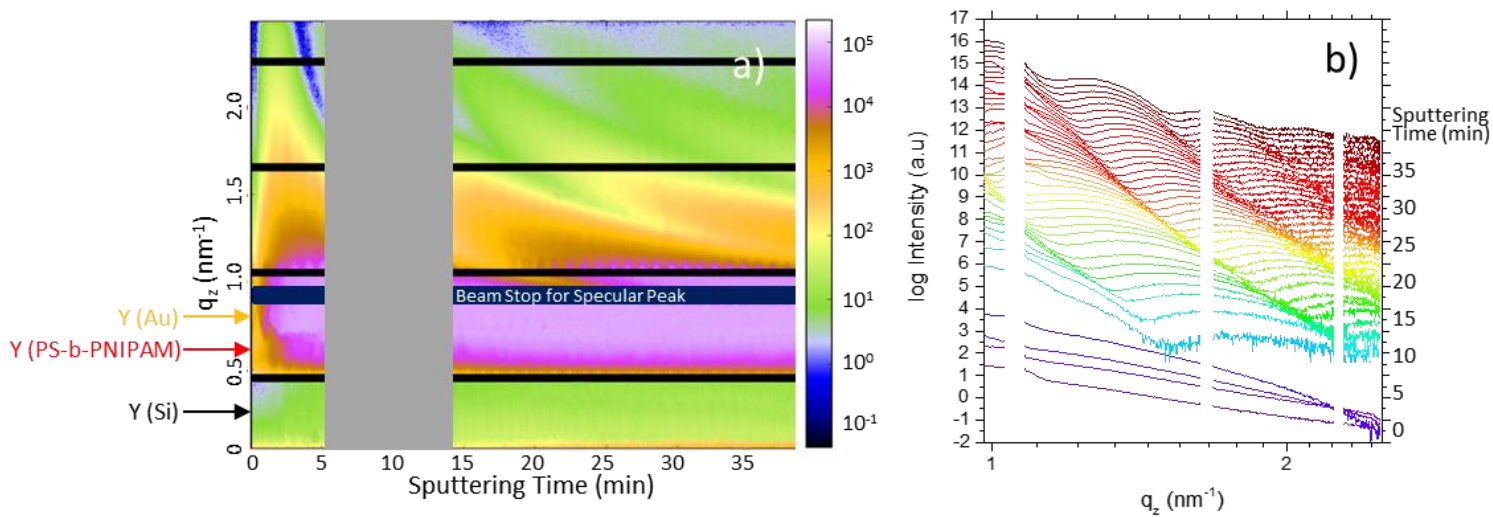

Figure 5-23: a) 2D time-resolved mapping showing the temporal intensity evolution of GISAXS vertical cuts at $q_y = 0$ nm$^{-1}$. The Yoneda peak positons of Au, PS-b-PNIPAM and Si are marked on the left. b) Intensity evolution of GISAXS vertical cuts at $q_y = 0$ nm$^{-1}$ with stacking on intensity scale for comparison.

From the vertical cuts of GISAXS profiles the layer information can be acquired and in figure 5-23, the 2D time-resolved mapping shows the intensity evolution of the vertical cuts at $q_y = 0$ nm$^{-1}$. The Yoneda peak positions of Au, PS-b-PNIPAM and Si are also indicated on the map. Since the sputtering chamber gives the constantly sputtering rate of 0.007 nm/s for gold nanoparticles, thus the thickness of the gold layer can be calculated theoretically. After 40 min sputtering of gold nanoparticles, the nominal thickness of gold layer is 16.8 nm.

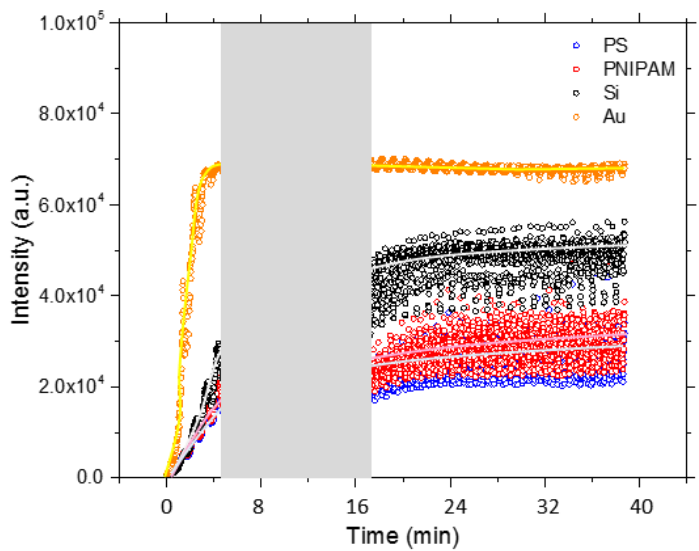

Figure 5-24: Evolution of Yoneda intensities versus sputtering time for PS (blue), PNIPAM(red), silicon (black) and gold (yellow).

The figure 5-24 presents the evolution of PS, PNIPAM, Si and Au Yoneda peak intensities versus Au sputtering time. Due to ill-defined structure of the prepared DBC thin films (as indicated by absence of any initial polymer characteristic peak), it is not obvious any selectivity phenomenon of gold to either PS or PNIPAM block of the DBC thin film. The selective wetting of gold to one or other block for block copolymer film systems during metal sputtering has been previously investigated by E. Metwalli et al [46-48]. And it has been observed that a strong selectivity to PS domains of the nanostructure PS-based DBC. The growth of the gold NPs on the polymer film show a typical exponential behavior as those reported for homo and block copolymers indicating a metal growth mechanism including four main steps; nucleation, diffusion, aggregation/coalescence processes and boundary movement between NPs [46-47].

# Chapter 6

# Conclusion and Outlook

In this thesis, the thermoresponsive behavior of free standing bulky nanostructured PS-b-PNIPAM diblock copolymer (DBC) films in water vapor environment is investigated. As well, the structures and the magnetic properties of the iron oxide/DBC hybrid bulk thick films prepared via solution-casting method are studied. The corresponding iron oxide/DBC hybrid thin films prepared using spin-coating methods are also presented. Several lab-based structural characterization methods as well as large-facility scattering techniques were used, such as scanning electron microscopy (SEM), small angle X-ray scattering (SAXS), grazing incidence small-angle X-ray scattering (GISAXS). Additionally, in-situ SAXS experiments to investigate swelling behavior of DBC in water vapor and in-situ GISAXS study of gold deposition on DBC thin films were performed. Finally, the magnetic properties of some selected metal oxide/DBC hybrid bulk films were investigated using superconducting quantum interference device (SQUID).

.

Here, a special temperature and humidity controlling cell was designed and constructed to perform in-situ SAXS study of the thermoresponsive behavior of nanostructured PS-b-PNIPAM DBC free standing bulk film. Two different PS-b-PNIPAM DBCs were studied, one with major PNIPAM weight fraction and the other with major PS weight fraction. The in-situ SAXS study of PS-b-PNIPAM DBC free standing film (with major PNIPAM block) included three temperature cycles including temperature above (40ºC) and below (20ºC) the characteristic lower critical solution temperature (LCST) of PNIPAM block at different relative humidity 5, 70 and 90%. The results show that the relative humidity has significant effect on the thermoresponsive behavior of the DBC bulk film at temperature below the LCST. At high relative humidity (90%), the PS-b-PNIPAM DBC bulk film systematically swells in water vapor within several hours at 20ºC. Switching from high to low (5%) relative humidity at 20ºC, a significant deswelling behavior of the DBC film is observed, indicating a strong sensitivity of the system to humidity environment at temperatures below the LCST. This behavior to the best of our knowledge has not been investigated in previously reported studies on nanostructured thermoresponsive hydrogel "dry" systems.

A phase transition from lamellar/cylinder mixed structures of the DBC bulk film to mainly cylinder morphology was observed after several hours' exposure time to water vapor environment at 20ºC. The impact of temperature on the PS-b-PNIPAM DBC free standing film is expected for a thermoresponsive system. As temperature jump from 20ºC to 40ºC, the swelled DBC film with cylinder morphology quickly shrinks going through a reversible phase transition to the initial lamellar/cylinder mixed structure state. During the three temperature cycles, the film was quite reversible upon swelling/deswelling process showing a stable and consistent interesting behavior. For the PS-b-PNIPAM DBC bulk film (major PS block) the swelling/deswelling behavior of the PNIPAM domains was much dramatically limited due the confinement effect of the rigid glassy PS matrix.

Additionally, to gain information about how far the swelling of PS-b-PNIPAM DBCs can proceed and to achieve an equilibrium state, an ex situ SAXS experiments were carried on for both PS-b-PNIPAM and PS-b-PNIPAM bulk films after water vapor swelling for 1~2 weeks. The results show that the PS-b-PNIPAM bulk sample again swells much less than that of the PS-b-PNIPAM bulk sample. For the PNIPAM major DBC, an equilibrated structure has not yet achieved even after two weeks of water vapor exposure. Possible deterioration/deformation of the free standing DBC films upon long time water vapor exposure has also be taken into account.

The interesting behavior of this bare nanostructured DBC to swell/deswell upon simple water vapor exposure, meaning changing the inter-domain spacing would imply a potential optical/sensor related application in case of metal or metal oxide incorporation in one or other polymer domain. The ability to manipulate the periodic spacing between nanoscale particles within metal oxide NPs/DBC hybrid nanomaterials using thermoresponsive type DBCs seems achievable as indicated from the bare DBC study.

Thus the next step is the SAXS investigation of PS-b-PNIPAM / iron oxide nanocomposite free standing film with different concentration of iron oxide contents. The aim here first is to study the incorporation mechanism of iron oxide nanoparticles within the PS-b-PNIPAM DBCs, and the influence of different concentration of iron oxide contents on the PS-b-PNIPAM DBCs structure. Two ways to incorporate iron oxide within the DBC, either using low decomposition temperature iron salt or $Fe_2O_3$ NPs. Upon annealing at high temperature, the iron salt in the iron salt/DBC films decomposes and forms iron oxide NPs. The iron salt is expected to reside within the PNIPAM block due to polarity. The SAXS investigation indicated that possibility to incorporate iron salt up to the [Fe]/[NIPAM] molar ratio of 0.05. At low [Fe]/[NIPAM] ratio, an expansion of the lamellar structured PS-b-PNIPAM DBC was observed. At high ratios >0.1, ill-defined morphology was observed as indicated from feature-less scattering behavior of these samples, likely due to formation of large particle aggregates. On the other hand, for the $Fe_2O_3$ NPs/DBC hybrid samples, the employed PS-coated $Fe_2O_3$ NPs is assumed to accommodate within the PS domain due to chemical compatibility. Our results indicated that the large sized 10 nm $Fe_2O_3$ NPs used in this study is relatively large compared to the PS domain size to uniformly disperse within the DBC resulted in an ill-defined morphology. Base on the results of the swelling behavior of bare DBC in water vapor, the swelling/deswelling behavior of metal or metal oxide/DBC hybrid system where one block is thermoresponsive is an interesting research direction for further investigation.

The corresponding $Fe_2O_3$ NPs/DBC hybrid thin films were investigated by static GISAXS experiments at Elettra, Italy. Also the gold metal deposition on PS-b-PNIPAM DBC thin film was studied using in-situ GISAXS experiments during gold sputtering process at DESY, Hamburg.

The superparamagnetic properties of iron salt/DBC and $Fe_2O_3$ NPs/DBC hybrid bulk materials were collected using a sensitive SQUID magnetometer, and the samples were measured on a film plane at different temperatures with external magnetic field various from -700 mT to 700 mT. The results for both nanocomposites show the hysteresis loops at low temperature (2K) and it becomes narrow and near to zero at high temperature (300K). These are the evidences that both iron salt/PS-b-PNIPAM and iron oxide/PS-b-PNIPAM nanocomposites behave as superparamagnetic materials.

# Appendix

## A.1 Cell Assembly, Test and Measurement

### A.1.1 Cell Assembly in Ganesha SAXS Instrument

The temperature and humidity controlling cell needs to be assembled by the following procedures to make sure the whole system works well. In table A-1 the detail procedures and components are listed and described.

| List of Cell Assembly Procedures and Detail Components | | | | |
|---|---|---|---|---|
| **1.** | **Assembly of temperature and humidity controlling cell outside:** First the Kapton window[1.5] should be place on both sides of the cell[1.1] and the aluminum frameworks[1.6] with four screws are used to tight the Kapton window[1.5] to make sure there is no leaking of vapor. The sample holder[1.3] should be placed in the cell vertically, then close the cell using the cap and rubber ring[1.4] with four long screws on the top. Finally the six Festo Push-in fittings[1.7] on the side should be screwed tightly. | | | |
| 1.1 | Cell Body | 50mm×60mm×50mm, Al | ×1 | TUM E13 Workshop |
| 1.2 | Bottom Block | 90mm, Al, combined with cell body, | ×1 | TUM E13 Workshop |
| 1.3 | Sample Holder | 20mm×15mm×1.5mm, Al | ×10 | TUM E13 Workshop |
| 1.4 | Cell Cap + Rubber Ring | 10mm×50mm×50mm, Al | ×1 | TUM E13 Workshop |
| 1.5 | Kapton Window | 25mm×20mm, Polyimide | ×2 | Provided by Anatoly |
| 1.6 | Window Framework | 25mm×20mm, Al | ×2 | TUM E13 Workshop |
| 1.7 | Push-in Fitting | Size 6 | ×6 | Festo |
| 1.7 | Tube | Size 6, 5 m | ×1 | Festo |
| **2.** | **Assembly of Humidity Regulation System** The humidifier[2.1] should be placed on a stable platform, and the water reservoir should be filled before switch on the machine. And also the nitrogen supply should be plugged in using normal tubes. The sensor[2.2] is screwed in the cell and another side is connected to the backside 9 pin Sub-D connecter of the humidifier using the re-configured extension cables[2.3] through the in-vacuum SAXS chamber to outside. The cell is placed on the sample stage in SAXS instrument and the in-let and out-let of the humidifier is connected with the cell inner space by plug-in vapor tube[2.4]. The USB-RS232 Cable[2.6] is used to connect the laptop controller software[2.5] and the humidifier. | | | |
| 2.1 | Humidifier | MHG 100 | ×1 | ProUmid |

| 2.2 | Humidity/Temperature Sensor | HC2-C04<br>Analog Out 1: 0…1V = 0…100%rH<br>Analog OUT2: 0…1V = -40…60°C | ×1 | Rotronic |
|---|---|---|---|---|
| 2.3 | Extension Cable | Adapted with HC2-C04 Sensor<br>Re-configured as shown in figure 5-3 | ×1 | Rotronic |
| 2.4 | Vapor Tube | Teflon, 3 meters | ×1 | TUM E13 |
| 2.5 | Controller Software | MHG Controller Software on Laptop | ×1 | ProUmid |
| 2.6 | USB-RS232 Cable | Cross-linked USB-RS232 Cable<br>Connected with Laptop | ×1 | Dr. Boehmer from Electronic Lager |
| 2.7 | Nitrogen Gas | 2-6 bar | ×1 | TUM Physics |
| **3.** | **Assembly of Temperature Regulation System**<br>The Julabo system[3.1] is used for the temperature regulation by water cycling. And normal tubes[1.7] and special tube connectors[3.2] are used to connect the Julabo water cycling system and the cell. Then put the temperature sensor[3.3] in the side hole of the cell for temperature feedback of Julabo system. | | | |
| 3.1 | Water Cycling System | Precisely temperature controlled water basin and cycling system | ×1 | Julabo |
| 3.2 | Tube Connector | In the SAXS instrument, prevents water leaking when disconnecting tubes | ×2 | Julabo |
| 3.3 | Temperature Sensor | In the SAXS instrument, connected with Julabo system, diameter 9 mm | ×1 | Julabo |
| **4.** | **Ready for the Test**<br>The whole system now is ready for the test before measurement. | | | |

Table A-1: List of assembly procedures and detail components for temperature and humidity cell.

## A.1.2 Cell Test and Measurement

The temperature and humidity controlling cell needs to be tested and the SAXS instrument also needs to be configured and aligned before measurement. Specific orders to stop the experiment should be kept. The detail procedures, commands and components are listed and described in table A-2.

## Table A-2: List of Cell Test Procedures and SAXS Configuration and Alignment Commands

**1. Cell tests needs to be done before measurement:**

Due to the operation of water cycling and vapor in the vacuum chamber, the cell must need to be tested before directly running the measurement.

| 1.1 | Screw Check | All the screws on the cell should be tighten especially the screws for the Kapton windows. | | |
|---|---|---|---|---|
| 1.2 | Plug Check | Check all the push-in plugs and make sure the tube cross section is smooth. | | |
| 1.3 | Vacuum Test | Put the cell inside the SAXS chamber and vent the system to check the stability of the cell in the vacuum system. The vacuum should be$< 10^{-2}$ mbar. | | |
| 1.4 | Water Leaking Check | Check the water leaking outside first using Julabo water cycling command. | | |
| | | Then check the water leaking in-vacuum by venting the SAXS chamber and running the Julabo water cycling. The vacuum should be$< 10^{-1}$ mbar. | | |
| 1.5 | Sample Position Check | Check if anything blocks the movement of the sample stage and if the sample can be positioned in and out the beam path for beam-stop overlapping by moving sample stage using command. | | |
| 1.6 | System Stability Test | After done all the check steps, leave the whole system running for 1 hour and observe the stability of the whole system. | | |

**2. SAXS instrument configuration and alignment of beam-stop, detector and sample**

Before the measurement, the SAXS instrument should be configured with right sample stage, beam-stop and detector distance. And the beam-stop, detector and sample need to be aligned also before measurement.

| 2.1 | Evacuate Chamber | evacuate_system | ×1 | Command window |
|---|---|---|---|---|
| 2.2 | Ramp X-ray | x_ramp ( to 49.8keV) | ×1 | Command window |
| 2.3 | Configure Sample Stage | conf_sample_stage 1 (106.8) | ×1 | Command window |
| 2.4 | Change beam-stop | conf_bstop 1 | ×1 | Command window |
| 2.5 | Detector Distance | conf_ugo 3 (SAXS);conf_ugo 1 (WAXS) | ×1 | Command window |
| 2.6 | Move sample out of beam path | mvr ysam {relative value}  or mv ysam {absolute value}, same for zsam. | ×1 | Command window |
| 2.7 | Move beam-stop to beam center | mv_beam2bstop | ×3 | Command window |
| 2.8 | Move sample back to the beam path | mvr ysam {value} mvr zsam {value} | ×1 | Command window |
| 2.9 | Find sample center | pd_in; o_shut; lup ysam -4 4 20 1 (same for zsam) | ×1 | Command window |
| 2.10 | Measurement Test | saxsmeasure 10 (check image) | ×1 | Command window |

**3. Temperature and Humidity Regulation System**

| | | Before the in-situ measurement, the temperature and humidity regulation should be on and running, and the humidifier should record the value of the humidity and temperature. | | |
|---|---|---|---|---|
| 3.1 | Temperature Regulation | Julabo_on<br><br>Julabo_stabilise 20 300<br><br>Julabo_off | ×1 | Julabo |
| 3.2 | Humidity Regulation | Switch on the humidifier, set the humidity value or run the method, and it is important to start log the file. | ×1 | ProUmid Software |
| 3.3 | SAXS Measurement | paste the edited macro and run it | ×1 | Command window |

**4.  Stop the Experiment**

To stop the running SAXS measurement, press ctrl + ^. And then type c_shut and vent_system in the command window. Stop the Julabo water cycling and switch off the humidifier, then open the chamber, disconnects all the tubes and cables inside the SAXS chamber. The vapor tube should be close from outside using a Festo connector. For the humidifier, the nitrogen gas should be stopped and the water in the reservoir should be evacuated. Keep the cell in the dry state and protect the Kapton window. Put the sensor in a plastic cylinder for protection.

| 4.1 | Stop Running SAXS Measurement | ctrl + ^ | ×1 | Command window |
|---|---|---|---|---|
| 4.2 | Close the Shutter | c_shut | ×1 | Command window |
| 4.3 | Vent the Chamber | vent_system | ×1 | Command window |
| 4.4 | Stop the Julabo System | Julabo_stop; Julabo_off | ×1 | Command window |
| 4.5 | Switch off the Humidifier | Switch off the button on the backside | ×1 | MHG Humidifier |
| 4.6 | Stop the Nitrogen Supply | Switch off the nitrogen valve | ×1 | Nitrogen Bottle |
| 4.7 | Disconnect tubes / cables | Open the chamber and disconnected all the tubes and cables, take out the Julabo sensor, then take out the cell | ×1 | SAXS Chamber |
| 4.8 | Close Vapor Tube | Close the vapor tube from outside using a Festo connector | ×1 | Festo |
| 4.9 | Storage of Humidifier | Remove all the connection cables and put it into standby mode, evacuate the water reservoir | ×1 | MHG Humidifier |
| 4.10 | Storage of Sensor | Put the sensor in a plastic cylinder | ×1 | Rotronic |
| 4.11 | Storage of Cell | Keep it in dry state and protect the Kapton windows. | ×1 | TUM E13 Workshop |
| 4.12 | Evacuate the Chamber | evacuate_system<br><br>Check the vacuum degree, should be less than $10^{-2}$ mbar | ×1 | SAXS Chamber |

## A.2 SEM Image of PS-b-PNIPAM DBC As-prepared Film

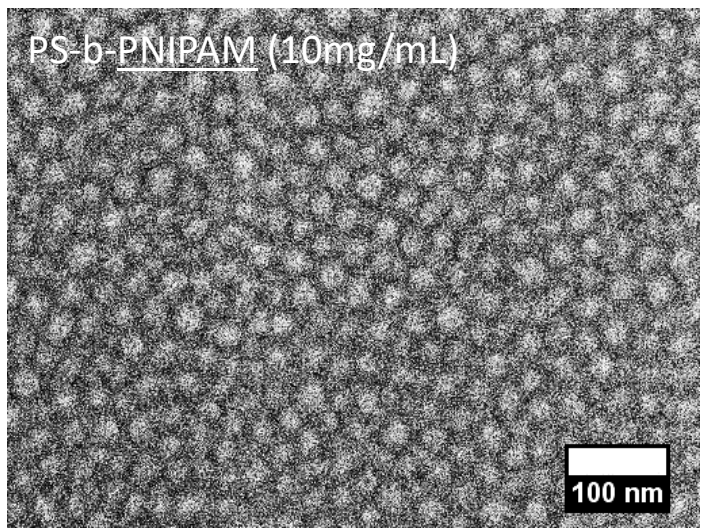

Figure A-1: SEM image bare PS-b-<u>PNIPAM</u> DBC as-prepared film with a low concentration of 10mg/mL.

# Acknowledgements

This Master Thesis was supported by Technical University of Munich and Ludwig Maximilian University of Munich under the framework of Erasmus Mundus Master MaMaSELF. I would like to thank my supervisor Prof. Dr. Peter Müller-Buschbaum from Physics Department at TU Munich and co-supervisor Prof. Dr. Wolfgang W. Schmahl from Crystallography Section at LMU Munich for their insights and expertise that greatly improved my master thesis. I also would like to thank Dr. Ezzeldin Metwalli as my advisor who provided great support and guidance during my thesis.

I express my warm thanks to Prof. Dr. Stephan V. Roth at DESY for the chance of in-situ sputtering GISAXS measurements, Dr. Sigrid Bernstorff at Sincrotrone Trieste for her great help during GISAXS/GIWAXS measurements, and Dr. Matthias Opel at Bavarian Academy of Sciences for SQUID measurements.

In addition, I would like to thank Senlin Xia, Thomas Kaps, Dan Yang, Max Kaeppel, Bernhard Springer, Dr. Konstantinos Raftopoulos, Bo Su, Lin Song, Weijia Wang from Chair E13 in Physics Department at TU Munich for their helps during my experiments and comments that greatly improved the manuscript, also thank Prof. Sergio Di Matteo, Prof. Philippe Rabiller and Mrs. Christiane Cloarec at Université de Rennes 1 and Mrs. Karin Kleinstück at TU Munich and LMU Munich for their continuous support, great help and suggestions during my master thesis.

Finally, an honorable mention goes to my family and relatives in Anqing China, and my friends Gang Luo in Beijing, Lijuan Zhang in Xiamen, Yifei Xu in Shanghai, Lan Peng in California, Zhiqin Liang in Toronto, Shuai Guo in Munich, Togzhan Nurmukanova in Rennes, Apoorva Ambarkar and Marta Mirolo in Grenoble for their understandings and supports on me during my Master Thesis.

Thank you all,

<div align="right">

Sept 30th 2016

Munich, Germany

</div>

Hong XU

Signature:

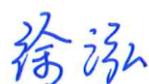

"I declare that I prepared and wrote this thesis work independently and with no other means than those referenced in the text".

Signature: Hong Xu

<div align="right">

30<sup>th</sup> Sept 2016
Munich, Germany

</div>

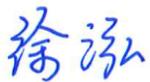